\documentclass[superscriptaddress,amsmath,pre,twocolumn,eqsecnum,a4paper]{revtex4-1}

\usepackage{graphicx}
\usepackage{amssymb}
\usepackage{bm}
\usepackage{hyperref}
\usepackage{ulem} % For \sout command, will not need this in the end
\makeatletter
\def\p@subsection{}
\def\p@subsubsection{}
\makeatother

\newcommand{\smallfig}{0.8\columnwidth} % figure width for "small" (narrow) figures
\newcommand{\largefig}{0.8\textwidth} % figure width for "large" (wide) figures

\newcommand*{\qav}[1]{\left \langle {#1} \right \rangle}

\newcommand{\fordraft}[1]{}
\newcommand{\toremove}[1]{}

\begin{document}
\normalem % For ulem package, to be removed in final version.

% For revtex4
\title{Non-adiabatic quantized charge pumping with tunable-barrier quantum dots: a review of current progress}
\author{Bernd Kaestner}
\affiliation{Physikalisch-Technische Bundesanstalt (PTB),
		Bundesallee 100, 38116 Braunschweig, Germany}
\email[Email:]{bernd.kaestner@ptb.de}
\author{Vyacheslavs Kashcheyevs}
\affiliation{Faculty of Physics and Mathematics,
		University of Latvia, LV-1002 Riga, Latvia}
\email[Email:]{slava@latnet.lv}

% For iopart:
%\article{Report on Progress}{Non-adiabatic quantized charge pumping with tunable-barrier quantum dots}
%\author{Bernd Kaestner}
%\address{Physikalisch-Technische Bundesanstalt (PTB),
%		Bundesallee 100, 38116 Braunschweig, Germany}
%\ead{bernd.kaestner@ptb.de}
%\author{Vyacheslavs Kashcheyevs}
%\address{Faculty of Physics and Mathematics,
%		University of Latvia, LV-1002 Riga, Latvia}
%\ead{slava@latnet.lv}

\begin{abstract}
%Charge pumping is a transport mechanism of generating electric current in the absence of a bias voltage. The pumping current results from periodic modulation of certain system parameters of a nanostructure connected to source and drain leads. Of particular interest has been the quantized regime when the current varies in steps of $e \cdot f$ as function of the system parameters, where $e$ is the electron charge and $f$ is the frequency of modulation. The adiabatic approach to quantized pumping relies on staying close to equilibrium during the modulation which has been explored extensively in the past in devices with tunnel barriers defined at the fabrication stage. In recent years, the field of precision quantized pumping has shifted towards a novel direction where non-adiabaticity is the underlying principle of operation. In contrast to adiabatic pumping this approach focuses on understanding of charge capture and release to and from periodically isolated regions, and the key for a successful implementation is the ability to employ tunable barriers with sufficiently wide dynamical range. Corresponding devices have been known since many years, but only recently experimental and theoretical approaches have converged towards a general picture, which is reviewed in the present manuscript. The precision of non-adiabatic pumping has approached the part-per-million level at frequencies several orders of magnitude higher than demonstrated so far in adiabatic pumps. Achievements towards envisaged applications will be discussed.
Precise manipulation of individual charge carriers in nanoelectronic circuits underpins practical applications of their most basic quantum property --- the universality and invariance of the elementary charge. A charge pump generates a net current from periodic external modulation of parameters controlling a nanostructure connected to source and drain leads; 
in the regime of quantized pumping the current varies in steps of $q_e f$ as function of control parameters, where $q_e$ is the electron charge and $f$ is the frequency of modulation. In recent years, robust and accurate quantized charge pumps have been developed based on semiconductor quantum dots with tunable tunnel barriers. These devices allow modulation of charge exchange rates between the dot and the leads over many orders of magnitude and enable trapping of a precise number of electrons far away from equilibrium with the leads. The corresponding non-adiabatic pumping protocols focus on understanding of separate parts of the pumping cycle associated with charge loading, capture and release. 
In this report we review realizations, models and metrology applications of quantized charge pumps based on tunable-barrier quantum dots.

%\\Submitted to \emph{Rep.\ Prog.\ Phys.} 
%but experimental and theoretical approaches have started to converge only recently. We present a tentative picture of this convergence, reviewing recent progress in implementations, modelling and applications of non-adiabatic quantized charge pumping with tunable barrier quantum dots.

%Slava: perhaps we still need this, but I don't see immediately a good place:
%The precision of non-adiabatic pumping has approached the part-per-million level at frequencies several orders of magnitude higher than demonstrated so far in adiabatic pumps. Achievements towards envisaged applications will be discussed.
\end{abstract}

% iopart-style:
% \submitto{\RPP}

% revtex4-style:
%\preprint{`Report of Progress' for Rep. Prog. Phys.}

% Comment out \maketitle if using iopart
\maketitle

\tableofcontents

\title{Non-adiabatic quantized charge pumping by tunable barrier quantum dots}


%
% Progress-Definition:
% We mean progress in understanding and application of non-adiabatic pumps.
% As for understanding - not only if a theory describes
% an experiment more precisely and makes predictions,
% but also if subsequently experiments show universality of the theory
% by demonstrating it in different material systems.
%

%% Calling in chapter files

%\input{introduction}
\section{Introduction}
\label{chap:Intro}

The ability to manipulate single charges provides access to various phenomena related to the quantization of electric charge~\cite{Averin1991, grabert1PBI} and  
has found important applications in the field of electrical metrology~\cite{keller1999, Flensberg1999a, Zimmerman2003, keller2007a}.
Corresponding devices are driven by alternating signals of a certain frequency $f$ in such a way, that an integer number $n$ of electron charges is \emph{pumped} through the device per cycle. The resulting \emph{quantized} current, $I = n \, q_e f$, can therefore be traced directly to the electron charge $q_e$ (we reserve $e =|q_e|$ to denote the fundamental constant of elementary charge). Besides their potential application as a single-electron based current standard they play an important role in the ongoing process of restructuring the International System of Units (SI)~\cite{mills2006, mise2009, milton2010a}. 
%In general, these devices can also be used to measure very low current levels with high precision, for which there is an increasing demand in science and industry [see BIPM key comparison database].
Although most of the recent developments discussed in the present Report have been focused around the metrological goals of fast and accurate single-electron delivery, the  few-electron on-demand sources built with metrology-inspired technology have already
found basic science applications in investigations of few-body mesoscopic physics and the development of tunable circuit elements for electron 
quantum optics \cite{FletcherPRL2013,FletcherPhysics2013,
NielsNature2014, Waldie2015}.

The scope and  organization of the Report is best illustrated
by explaining the meaning of the title.
Quantized pumping may be realized by employing an energy gap for removal/addition of single electrons to a small conductor~\cite{geerligs1990}, e.g.\ a metallic island or a semiconductor quantum dot (QD) in the Coulomb blockade regime. 
There is a trade-off in designing the tunnel barriers between the QD and leads: the conductance
should be  low enough to suppress charge fluctuations (and thus keep $n$ well-defined) but not too small to allow enough time for charge loading and unloading (and thus push up the operation frequency $f$ and hence the output current level). Hence the circuits based on \emph{fixed} tunnel barriers naturally lend themselves to the realization of \emph{adiabatic} pumping schemes where the number of confined charges is kept close  to equilibrium (defined by the leads and the environment) during most parts of the operation cycle.
A notable example of a successful implementation of the adiabatic approach is the development of single-electron-tunneling  pumps consisting of a series of small metallic islands~\cite{pothier1PBI} or, recently, atomic  donor states \cite{Roche2013} coupled by fixed tunnel barriers.
Using this principle Keller~\emph{et al.}~\cite{keller1996} achieved an uncertainty of $15 \times 10^{-9}$ with $f$ up to a few MHz determined by electron counting.
A comprehensive review of single-electron metrology with fixed-barrier single-electron transistors has been compiled by Flensberg \emph{et al.}~\cite{Flensberg1999a} in 1999 (see also Likharev~\cite{Likharev1999}).

In recent years, the field of accurate quantized current sources has shifted towards a  novel direction in which
the speed versus precision trade-off is optimized by employing  \emph{tunable barriers} with a sufficiently wide dynamical range (typically relying on field-induced conductance pinch-off in semiconductors, see Section 2.1 below). In such devices a precise number of electrons can be trapped on the QD at potential 
levels multiple charging energies away from the external electrochemical potential, making the pumping cycle strongly \emph{non-adiabatic} with respect to electron number equilibrium along the charge transfer path. 
Non-adiabatic  pumping relies 
on  switching charge exchange with external reservoirs on and off in a well-defined and temporally separate manner
and thus enables greater flexibility and simplicity in design and analysis of quantized charge transfer protocols compared to entirely adiabatic schemes. 
This admittedly narrow scope for defining non-adiabaticity is further explained in Sections \ref{sec:plungerBarrier} and \ref{sec:modIntro}; it should not be confused with inevitable deviations from strict instantaneous equilibrium due to irreversible excitations of gapless degrees of freedom.
At the time of writing optimized tunable barrier pumps operating in the GHz frequency range have been demonstrated to beat the uncertainty limit of state-of-the-art current-measurement setups~\footnote{Note that an evaluation by counting in the GHz regime has not yet been realized.}, which lies around $10^{-6}$ ~\cite{giblin2012}.

The purpose of the present Report on Progress is to bring together different elements of what we believe is an emerging
coherent picture of non-adiabatic charge pumps which are  based on semiconductor quantum dots with tunable barriers. 
There are three connected but relatively independent components to this story gathered in the main sections below.
Section~\ref{sec:real} reviews the basic design principles and the experimental state of the art in quantized pumping with tunable-barrier devices (turnstiles, adiabatic and non-adiabatic pumps, as well as devices employing surface acoustic waves  and Josephson junctions for barrier modulation).  
Section~\ref{chap:quantModel} singles out the relative simplicity, robustness and universality of
a particular capture-limited non-adiabatic pumping protocol -- the single-gate semiconductor quantum dot pump (introduced in the wider context of Section~\ref{sec:plungerBarrier}). Section~\ref{chap:quantModel} is organized according to the availability of compatible and comparable models, realizations and measurements.
The necessary theory elements are collected in Section~\ref{sec:modIntro} and put to use in Sections~\ref{sec:currModel}--\ref{sec:CountStatModel} where the widely-used  decay cascade model and its generalizations are described in the context of supporting experiments. A theoretically-minded reader might find Section~\ref{sec:modIntro} a useful starting point for connecting with a wider spectrum of experimental approaches from  Section~\ref{sec:real}, not necessarily restricted to single-parameter pumps. Finally,
Section~\ref{chap:metrology} approaches quantized current sources from the metrology perspective. It offers an overview of the present state of the art on the quantum metrological triangle, the progress and challenges for pump accuracy optimization, and an outlook for error accounting in a self-referenced realization of quantum ampere.

Readers interested in  other types of single-electron sources and their basic physics applications are encouraged to consult a recent review article by Pekola~\emph{et al.}~\cite{Pekola2013}. A review of non-equilibrium coherent phenomena in single-electron quantum optics has been created by Bocquillon \emph{el al.}~\cite{Bocquillon2014}.  A more applied perspective focusing on single-electron-based-circuits can be found in the review by Ono~\emph{et al.}~\cite{ono2005}.

%%%%%%%%%%%%%%%%%%%%%%%%%%%%%%%%%%%

%\input{realizations}
%%%%%%%%%%%%%%%%%%%%%%%%%%%%%%%%%%%%%%%%%%%%%%%%%%%%%%%%%%%%%%%%%%%%%%%%%%%%%%%%%%%%%%%%%%%%%%%%
%%%%%%%%%%%%%%%%%%%%%%%%%%%%%%%%%%%%%%%%%%%%%%%%%%%%%%%%%%%%%%%%%%%%%%%%%%%%%%%%%%%%%%%%%%%%%%%%
%%%%%%%%%%%%%%%%%%%%%%%%%%%%%%%%%%%%%%%%%%%%%%%%%%%%%%%%%%%%%%%%%%%%%%%%%%%%%%%%%%%%%%%%%%%%%%%%

\section{Overview of driven tunable-barrier devices}
\label{sec:real}
In this section we introduce some of the main ideas that underpin quantized current generation by electrostatic modulation of tunable-barrier quantum dot (QD) devices. The discussion is largely qualitative and is structured around experimentally demonstrated approaches. We briefly review the basics of tunable-barrier semiconductor devices (Section~\ref{sec:barriers}), Coulomb blockade and single-electron turnstiles (Section~\ref{sec:BarrierMod}), then introduce the concept of non-adiabatic quantized pumping (Section~\ref{sec:plungerBarrier}) which is the core topic of our Report. This section is supplemented by an overview of tunable Josephson junction devices (Section~\ref{sec:Joseph}) and surface-acoustic-wave-induced single-electron pumping in Section~\ref{sec:SAW}.

%%%%%%%%%%%%%%%%%%%%%%%%%%%%%%%%%%%%%%%%%%%%%%%%%%%%%%%%%%%%%%%%%%%%%%%%%%%%%%%%%%%%%%%%%%%%%%%%
\subsection{Tunable barriers in semiconductors}
\label{sec:barriers}

The semiconductor device structures and processing techniques forming the basis of tunable barrier QDs are described in textbooks by, e.g., S. M. Sze~\cite{sze1PAI} or R. Williams~\cite{williams1POI}. The resulting devices may be considered as special realizations of field effect transistors (FETs), also known e.g. as HEMT (High Electron Mobility Transistor), MODFET (MOdulation Doped Field-Effect Transistor) or MOSFET (Metal-Oxide-Semiconductor Field-Effect Transistor)~\cite{sze1PAI}.

\begin{figure}
\includegraphics[width=\smallfig]{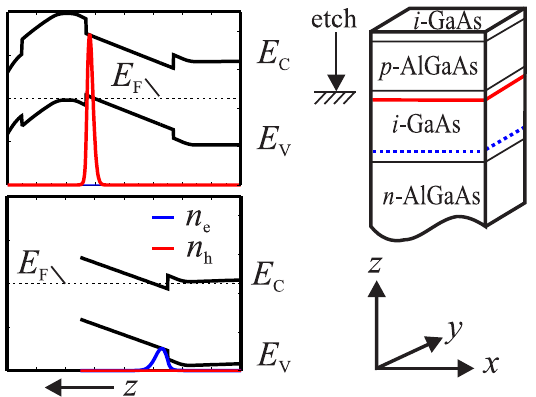}
\caption{Example of an energy band diagram ($E_{V,C}$) and corresponding electron and hole density ($n_e$, $n_h$) of $n$- and $p$- doped AlGaAs in contact with undoped GaAs. The substrate surface is to the left of the diagram. The lower diagram shows a possible result of etching, where it removes holes from the GaAs quantum well (QW) while $n_e$ is increased.
\label{fig:barrier}}
\end{figure}

A starting point is, for example, a substrate with a thin layer of charge carriers in the vicinity of the surface. This can be achieved by ion implantation or incorporating a thin sheet of dopands during growth.
%, resulting in so called $\delta$-doping~\cite{tsukagoshi3PBI}.
Conducting layers with special two-dimensional transport properties~\cite{ando1PBII, bastard1PBII, Datta1995, davies1997} have been created in \emph{heterostructures} where the different layers of material have unequal bandgaps. 
An example of a heterostructure combining AlGaAs and GaAs is shown Figure~\ref{fig:barrier} along with a graph tracing the corresponding conduction ($E_C$) and valence band ($E_V$) edges as function of the coordinate perpendicular to the layers.  Doping profiles are designed to shift band edges with respect to the Fermi level ($E_F$) in order to populate the interface regions, also called quantum wells (QW), with charge carriers. Details on this charge transfer process and corresponding band diagram profiles can be found in textbooks, such as by G. Bastard~\cite{bastard1PBII}. Depending on the carrier type, the charge carriers are described as two-dimensional electron (2DEG) or hole gas (2DHG).

\begin{figure*}
\includegraphics[width=\largefig]{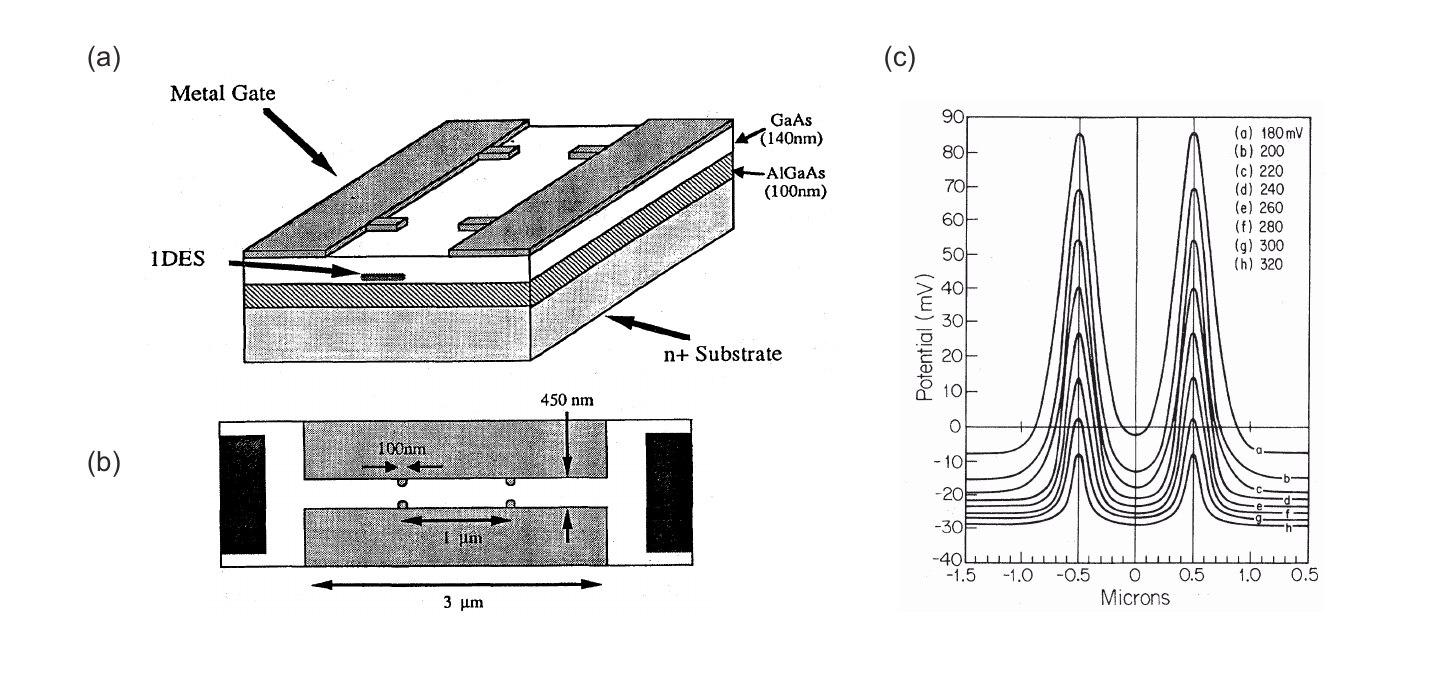}
\caption{Schematic drawings of a structure to confine charges in all three dimensions~\cite{kastner2PBI}. (a) A one-dimensional electron system forms at the top of the GaAs-AlGaAs interface, with density controlled by the substrate voltage $V_g$. Panel (b) shows the dimensions used for the potential calculation in (c). Reprinted with permission from~\cite{kastner2PBI}. Copyright 1992 by the American Physical Society.}
\label{fig:barrier_lat}
\end{figure*}

A wire-like geometry is defined by imposing a corresponding electrostatic environment or permanent surface modifications. The latter can be achieved by, e.g., ion bombardment or etching~\cite{williams1POI}. A deep etch would remove the QW completely, while in a shallow etch the resulting close proximity of the QW to the surface causes depletion of carriers from it. Note that a high density of surface states may lead to pinning of $E_F$ at the surface in some materials. The doping profile can also be engineered so that a shallow surface etch causes a change of carrier type~\cite{kaestner2007, wunderlich}, as shown in Figure~\ref{fig:barrier}.

Tunable barriers along the wire can be achieved by imposing spatially varying electrostatic potentials shifting the energy band locally. Such electrostatic environments can be provided by, e.g., depositing gates on the surface, which are galvanically separated from the transport channel by Schottky contact formation or dielectrics~\cite{sze1PAI}. Another technique defines conducting regions out of the 2DEG or 2DHG acting as in-plane gates. Transport and gate regions may be separated via standard etch techniques, or other methods specifically designed for nanoscale structures~\cite{snow1995a, schumacher2000, crook2003}. The cool-down procedure itself may also influence the nanoscale electronic properties of heterostructures at low temperatures due to different frozen charge configurations on impurities and defects~\cite{Pioro2005}. 

The above techniques allow to confine charge carriers in all three dimensions which may result in the definition of QDs.
A particular example of a tunable-barrier QD structure is shown in Figure~\ref{fig:barrier_lat}(a)~\cite{kastner2PBI}. A positive voltage $V_g$ applied to the heavily doped $n^+$ substrate controls the electron density. Negatively biased metal gates on the surface not only confine the carriers along a narrow wire but also form constrictions at $1\,\mu$m distance along the transport channel. Lowering the backgate voltage $V_g$
the electron density under the gates is reduced, which leads to a corresponding decrease in conduction and tunneling. The calculated variation 
of the conduction band edge is plotted as function of position along the channel in Figure~\ref{fig:barrier_lat}(c). At $V_g = 180\,$mV the minimum in the potential between the barriers drops just below the Fermi energy (zero level in the figure). This results in electron accumulation in this region. Further increase of the gate voltage lowers the barriers, so that at about $300\,$mV the QD definition is lost. The potential landscape and actual transmission through this type of barrier has been investigated intensively~\cite{Laux1988, Wu1993, Davies1995, Sun1995, kristensen2000}, motivated largely by need to understand mesoscopic effects seen in conducting quasi-one-dimensional channels such as conductance quantization~\cite{Imry1999}. 

The potential barriers in the devices studied in this Report are typically tuned by individual gates.  In Section~\ref{chap:quantModel} we discuss the corresponding theoretical approaches and the implications of exponential barrier tunability for clocked electron transfer.

%%%%%%%%%%%%%%%%%%%%%%%%%%%%%%%%%%%%%%%%%%%%%%%%%%%%%%%%%%%%%%%%%%%%%%%%%%%%%%%%%%%%%%%%%%%%%%%%
\subsection{Tunable-barrier turnstiles}
\label{sec:BarrierMod}

\begin{figure}
\includegraphics[width=\smallfig]{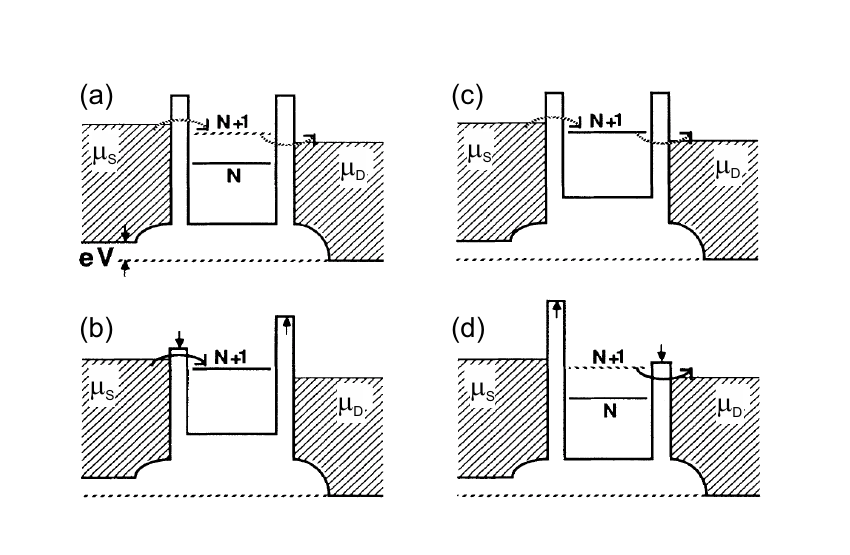}
\caption{Schematic potential landscape sequence for turnstile operation~\cite{kouwenhoven1PBI}. The electrochemical potentials of the source and the drain reservoirs are denoted by $\mu_S$ and $\mu_D$, respectively. The level $N$ between the potential barriers denotes the electrochemical potential $\mu_N$ of the QD with $N$ electrons in it. Adapted with permission from~\cite{kouwenhoven1PBI}. Copyright 1991 by the American Physical Society.}
\label{fig:Kouw0}
\end{figure}

One of the simplest quantized current generation schemes for a tunable-barrier quantum dot is a single-dot \emph{turnstile}, proposed by Odintsov~\cite{odintsov1PBI} and realized
in a pioneering work by Kouwenhoven~\emph{et al.}~\cite{kouwenhoven1PBI, kouwenhovenB91}. 
The current in a turnstile is driven by an external dc  bias while the clocked switching of the barriers on and off ensures the desired order of electron transfer events, as illustrated in Figure~\ref{fig:Kouw0}. In order to obtain quantized current the number of tunneling events has to be controlled each time, which is achieved by employing Coulomb blockade of tunneling~\cite{Fulton1987, Averin1991, grabert1PBI}.

The basic principle of Coulomb blockade can be qualitatively understood by referring to Figure~\ref{fig:Kouw0} adopted from \cite{kouwenhoven1PBI}. It shows schematically the potential landscapes of the QD connected to the leads. The electron states in the source (S) and drain (D) reservoirs are occupied up to the electrochemical potentials $\mu_S$ and $\mu_D$, respectively, which differ due to the bias voltage $V= (\mu_S - \mu_D)/q_e$. The line labeled $N$ denotes the electrochemical potential $\mu_N$ of the QD when it contains $N$ electrons. Addition of an extra electron to the QD into the lowest available energy state would increase the electrochemical potential to $\mu_{N+1}$, indicated by line $N+1$ in the figure. Further electron tunneling into the QD will be suppressed if $\mu_S, \mu_D < \mu_{N+2}$ (Coulomb blockade). The addition energies $\mu_{N+1}-\mu_{N}$ for the relatively large GaAs QDs employed in~\cite{kouwenhoven1PBI} are dominated by the capacitive charging energy $E_c=e^2/C$ where $C$ is the sum of the capacitances between the QD and the different gates. 
Coulomb blockade with single-electron resolution requires a sufficiently low temperature such that $k T \ll E_c$, with $k$ Boltzmann's constant. Hence the measurements are typically performed at cryogenic temperatures. 
Additionally, suppressing the quantum uncertainty of the electron number on the QD requires the product of the charging energy 
$e^2/C$ and the tunnel-barrier $RC$-time to be smaller than the Planck constant $h$, hence the conductance between each of the leads and the dot, $G$, must  remain smaller than the conductance quantum, $e^2/h \approx (12.9 \, \rm{k}\Omega)^{-1}$, for the Coulomb blockade to hold.

%Control over the number of tunnel events can be gained by keeping the thermal energy of the electrons small compared to the addition energy, $kT \ll e^2/C
%$, allowing 

\begin{figure}
\includegraphics[width=\smallfig]{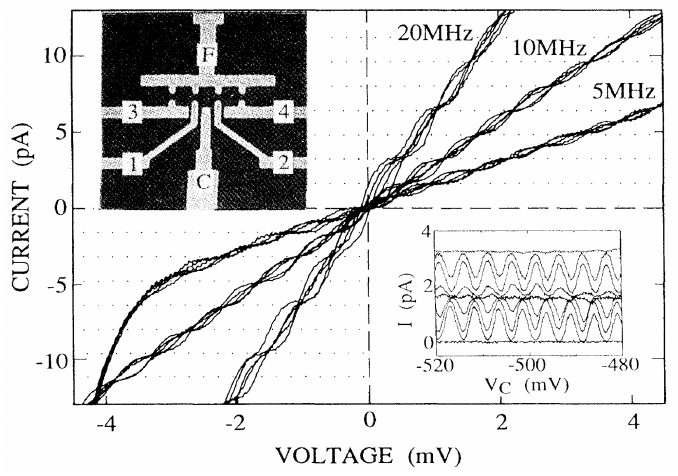}
\caption{$I$-$V$ curves of the QD driven at frequencies $f=5, 10,$ and $20\,$MHz for different voltages applied to gate C~\cite{kouwenhoven1PBI}. Dotted lines indicate multiples of $q_ef$ for $f=10\,$MHz. Upper inset: Image of the device with gate layout forming quantum point contacts (QPCs) at the constrictions. Lower inset: Current versus voltage on gate C  for $f=10\,$MHz and different fixed bias voltages. Reprinted with permission from~\cite{kouwenhoven1PBI}. Copyright 1991 by the American Physical Society.}
\label{fig:Kouw}
\end{figure}

The turnstile sequence of operation for a Coulomb-blockaded QD is shown in Figure~\ref{fig:Kouw0}. 
%In (a) and (c) the QD is tuned into a regime which allows sequential tunneling of an electron on and off the QD, but with high tunnel resistances (dashed arrows). 
The potential landscapes in Figure~\ref{fig:Kouw0}(b) and (d) represent the loading and the unloading phases of the cycle, during which the QD can equilibrate with the source and the drain reservoir, respectively, owing to the low height of the corresponding barriers (see solid arrows). Between these phases the barriers are sufficiently high (see dashed arrows) so that the charge on the QD will remain stable on the timescale of a cycle keeping either $N$ (phase (a) in Figure~\ref{fig:Kouw0}) or $N+1$ (phase (c) in Figure~\ref{fig:Kouw0}) electrons. The potential sequence from (a) to (d) causes the integer difference $n$ in the number of electrons acquired from the source and the drain, respectively, to be transported through the QD. For the scheme depicted in Figure~\ref{fig:Kouw0} this number is $n=(N+1)-N=1$, but $n$ can be increased by increasing the bias voltage and hence the number of charge states within the energy interval between $\mu_S$ and $\mu_D$,  yielding a quantized current, $I = n \, q_e f$, where $f$ is the repetition frequency. The turnstile operation manifests itself as current plateaus in the $I$-$V$ characteristic corresponding to integer multiples of $q_e f$ as shown in Figure~\ref{fig:Kouw} for different frequencies $f$.

The device used by Kouwenhoven~\emph{et al.} has been realized in the 2DEG of a GaAs-AlGaAs heterostructure, shown in the inset of Figure~\ref{fig:Kouw}. The QD is defined between gates labeled 1, 2, C and F. Here gates 1 and 2 define the barriers and gate C acts as plunger (gates 3 and 4 are grounded). The small total capacitance of $C= 240\,$aF of the QD ensures a controllable discrete number of charge on the QD at the measurement temperature of 10$\,$mK. The two voltage signals on gate 1 and 2 are modulated with frequencies up to 20$\,$MHz and with a $180^\circ$-phase shift, hence modulating the barriers to the left and right reservoir.
% QDs discussed in this section show a finite cross-talk between the barrier tuning
%gate voltages and the electrochemical potential of the container.
For optimal realization of the turnstile scheme the barriers should be modulated independent of the island potential, which ideally should be kept fixed throughout the cycle. 
%\sout{In real QD devices there will be cross-talk between the plunger and barrier gates.} 
Compensation of the barrier-plunger cross-coupling has been achieved in this scheme employing $180^\circ$-shifted harmonic signals and ensuring the similarity of the capacitances between the barrier-defining gates 1,2 and the QD.

\begin{figure}
\includegraphics{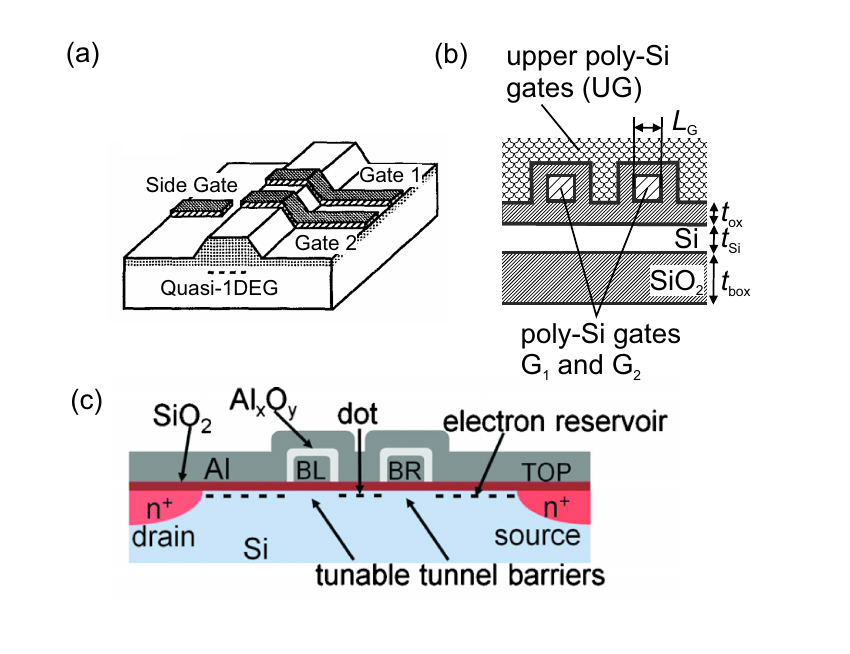}
\caption{(a) Device type used by Nagamune~\emph{et al.}~\cite{nagamune1}. The wire is 460nm wide, but due to the depletion of electrons at the edge the conducting region is roughly 400$\,$nm. Gates are 230$\,$nm wide and separatred by a gap of 330$\,$nm. Reproduced with permission from~\cite{nagamune1}. Copyright 1994, AIP Publishing LLC. (b) Schematic cross section of the device type used by Fujiwara~\emph{et al.}~\cite{fujiwara1}. The thickness of the wire ($t_\mathrm{Si}$), the gate oxide ($t_\mathrm{ox}$) and the buried oxide ($t_\mathrm{box}$) are about 20, 30 and 400nm, respectively. Reproduced with permission from~\cite{fujiwara1}. Copyright 2004, AIP Publishing LLC. (c) Schematic cross section of the device used by Chan \emph{et al.}~\cite{Chan2011a}. Two aluminium barrier gates (BL and BR) are crossed by a top gate (TOP) isolated by Al$_x$O$_y$. Reproduced with permission from~\cite{Chan2011a}. Copyright 2011, AIP Publishing LLC.}
\label{fig:BarrMod_Devs}
\end{figure}

The device structure used by Nagamune~\emph{et al.}~\cite{nagamune1} for turnstile operation was realized by etching narrow wires in a GaAs-AlGaAs heterostructure, as shown in Figure~\ref{fig:BarrMod_Devs}(a). This leads to an increase in the charging energy due to a smaller size of the dot and a reduced amount of metal that screens intradot Coulomb interaction. 
Clear current quantization at a measurement temperature of $10\,$mK has been observed (taking into account the effect of  an additional conductance path).

A prominent feature of single-electron, single-QD turnstiles realized in \emph{silicon} is that fabrication is more suitable for smaller feature sizes and consequently operation at a relatively high temperature. Ono~\emph{et al.}~\cite{ono2003} realized turnstile operation in the low MHz range using closely spaced MOSFETs at a temperature of 25$\,$K. Further studies were carried out by Fujiwara~\emph{et al.}~\cite{fujiwara1} and their device consisted of a 30$\,$nm wide silicon nanowire crossed by poly-Si gates of $L_G = 40\,$nm length. In addition there is an upper poly-Si gate as shown in Fig.~\ref{fig:BarrMod_Devs}(b). Because the upper gate intrudes into the gap between the fine gates, the QD is controlled by the upper gate in a self-aligned way. The total capacitance is estimated to be of the order of 10$\,$aF. Turnstile operation up to 100$\,$MHz was investigated at T = 20$\,$K. The error was estimated to be below 10$^{-2}$ at 100$\,$MHz. The device could be tuned into a regime where all electrons captured from source move to the drain so that the island is completely depleted periodically. The number of electrons transferred can thus be entirely controlled by the upper-gate voltage (plunger gate). Related turnstile devices have been studied by Yamahata \emph{et al.} \cite{yamahata2011,YamahataPRB2014}, which will be discussed below in Section~\ref{sec:CountStatModel} in the context of error rate evaluation.

A different  silicon-based QD system has been employed by Chan~\emph{et al.}~\cite{Chan2011a}. The device was fabricated on a high resistivity silicon substrate and the conducting layer is induced at the Si-SiO$_2$ interface using positive gate voltages on the top gate. The schematic cross-section is shown in Fig.~\ref{fig:BarrMod_Devs}(c). Measurements were carried out at 300$\,$mK and plateaus were observed up to frequencies of 240$\,$MHz. Simulation of the observed results within a sequential tunneling model with exponentially tunable rates (see Section \ref{sec:modIntro} below) suggests substantially elevated effective temperature, attributed in \cite{Chan2011a} to heating of the electron gas in the source and drain electrodes by the ac driving voltage.

%%%%%%%%%%%%%%%%%%%%%%%%%%%%%%%%%%%%%%%%%%%%%%%%%%%%%%%%%%%%%%%%%%%%%%%%%%%%%%%%%%%%%%%%%%%%%%%%
\subsection{Tunable-barrier pumps}
\label{sec:plungerBarrier}

\begin{figure}
\includegraphics[width=\smallfig]{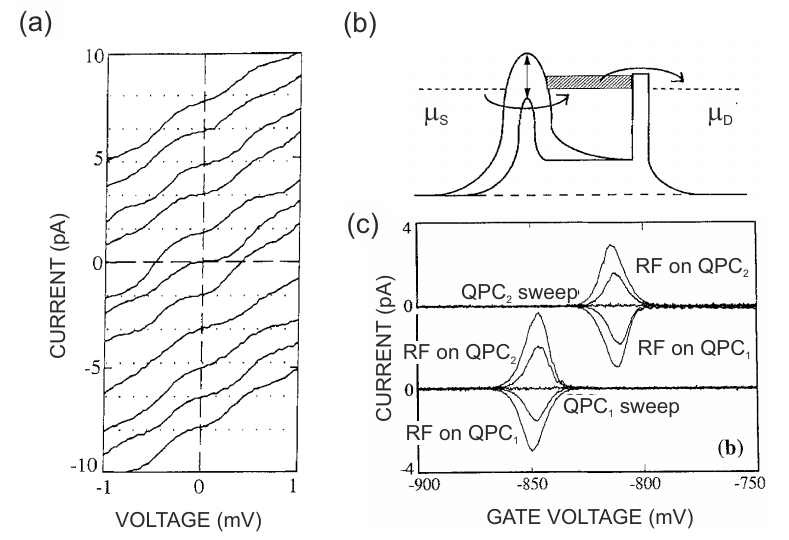}
\caption{(a) $I-V$ characteristic for a similar device as shown in Figure~\ref{fig:Kouw} with settings for voltage on gate C, rf amplitudes and phase differences such that current is quantized at zero bias voltage. (b) Schematic potential landscape for electron pumping using one barrier. (c) Non-quantized pumping current for different gate voltage settings for a similar device as presented in Figure~\ref{fig:Kouw}. Reproduced with permission of the authors of~\cite{kouwenhovenB91} and Springer Science+Business Media.}
\label{fig:Kouw2}
\end{figure}

The simple turnstile mechanism depicted in Figure~\ref{fig:Kouw0} requires the energy of the $(N+1)$-th electron to be confined to the bias window, 
$\mu_S > \mu_{N+1} >\mu_D$. This is challenging for large-amplitude modulation because of crosstalk from the barrier-defining gate to the potential on the QD. Instead of electrostatic compensation of the level movement during the turnstile operation (e.g, by left-right symmetric design and modulation \cite{kouwenhoven1PBI} as discussed in the previous section)  one can utilize both the barrier and the plunger functions of the gates to operate a charge transfer scheme without the external  voltage bias. A device producing directed current output under periodic driving for equal source and the drain potentials is commonly called a charge \emph{pump} \cite{pothier1PBI}. The same device as shown in Figure~\ref{fig:Kouw} and described as a turnstile in the previous section can act as a pump if the right kind of asymmetry in modulation amplitude and phase is applied to the barrier-creating gates. This has been demonstrated in \cite{kouwenhovenB91}, and an example is shown in Figure~\ref{fig:Kouw2}(a). A series of $I$-$V$ curves are shown for the same device as in Figure~\ref{fig:Kouw}  where
the modulation  amplitudes on the barrier gates 1 and 2,  the relative phase as well as the voltage on the plunger gate C have been adjusted so that plateaus from $-5 \, q_ef$ to $+ 5 \, q_ef$ appear around zero bias voltage. The measurement was carried out at a temperature of 10$\,$mK and the pump frequency set to 10$\,$MHz. Despite the relatively poor quantization these findings have shown a possible route to harness the cross-capacitances which become increasingly important for smaller QD feature sizes (we discuss a measure of relevance for this barrier-plunger crosstalk in Section~\ref{sec:modIntro}).  

\begin{figure}
\includegraphics[width=\smallfig]{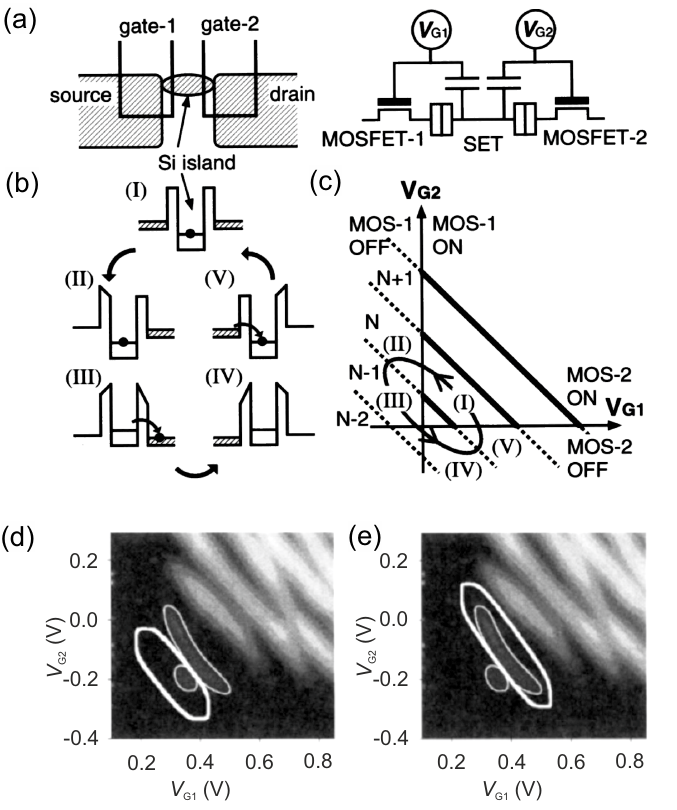}
\caption{Double-barrier single electron pump realized in silicon by Ono and Takahashi~\cite{Ono2003a}. (a) shows the lithographic geometry and the circuit diagram. (b) illustrates the pump cycle in terms of energy diagrams, and in (c) within the 2D conduction map of the MOSFET defined QD structure. Conduction maps measured at 25$\,$K are shown in (d) and (e), together with the contours of the first and  the second conduction peak indicated by thin white loops. Two sample trajectories for $q_e f$ current generation are shown by the thick white loops in the two images (d) and (e). Reproduced with permission from~\cite{Ono2003a}. Copyright 2003, AIP Publishing LLC.
\label{fig:ono}}
\end{figure}

A double-barrier single-electron pump has been realized with silicon-based MOSFETs by Ono and Takahashi~\cite{Ono2003a}. They have carried out a systematic study relating the pump current to the dc conductance as function of both barrier gate voltages,
as shown schematically in Figure~\ref{fig:ono} (a-c).  The gates where designed to control efficiently both the conductance of the MOSFETs \emph{and} the electrostatic potential of the middle island (the QD), see Figure~\ref{fig:ono} (a). Figure~\ref{fig:ono}(b) illustrates the single-electron pumping protocol implemented in \cite{Ono2003a} which does not require source-drain bias across the pump. State I represents a Coulomb-blockaded state of the QD containing $N$ of electrons. Closing the left channel by applying a negative bias to gate $1$ leads to state II. The island potential is kept nearly constant by applying a positive control bias to gate $2$. In order to eject the electron to the right channel the island potential is raised ending up with a new Coulomb blockade state with $N-1$ electrons (state III). State IV is reached by opening the left channel and simultaneously closing the right channel, 
which keeps the island potential nearly constant. The cycle finally enters state V by lowering the island potential so that an electron can enter from the left channel. 

A map of dc conductance as function of both barrier voltages reveals the Coulomb resonances which correspond to a match in energy between  $\mu_S = \mu_D$ and the electrochemical potential $\mu_N$ for electron addition or removal between $N-1$ and $N$ electron states on the QD, thus allowing one to choose the optimal path in the $V_{g1}$-$V_{g2}$ plane for the realization of the quantized pumping sequence, see Figure~\ref{fig:ono}(c). The contour corresponding to the single-electron transfer protocol of  Figure~\ref{fig:ono}(b) can be found by encircling the maximum of resonant conduction as shown Figure~\ref{fig:ono}(d) and (e). The pump current was measured at a temperature of 25$\,$K for a range of frequencies up to 1$\,$MHz. A measurement-limited uncertainty of the order of $10^{-2}$ was obtained. 

Detailed experimental studies of quantized pumping with contours encircling the resonance lines have been performed by Jehl~\emph{et al.}~\cite{jehl2012, jehl2013} using a metallic NiSi nanowire-island system with self-aligned MOSFETs, fully integrated into an industrial microelectronics process. The gate design enabled a significant increase in operation frequency up to 1$\,$GHz. At a measurement temperature of $0.6\,$K and encircling $N=7$ resonance lines quantized currents of 1.12$\,$nA have been generated --- a level which is metrologically relevant (see Section~\ref{sec:metro}).

An important feature of the pumping cycle shown in Figure~\ref{fig:ono}(b) is that 
it is composed of a sequence of \emph{equilibrium} states and the desired operation relies on the device following this sequence of states sufficiently closely, i.e.\ \emph{adiabatically}. 
Average charge per cycle transferred between the source and the drain by an adiabatic pump \cite{pothier1PBI} 
is determined solely by the equilibrium charge diagram and the pumping contour, but not the rate at which the contour is  traversed.
The connection between resonant transmission and quantized charge pumping has been studied theoretically~\cite{Levinson02PhA,Entin02res,Kashcheyevs2004,Fioretto08} in the context of adiabatic quantum pumping~\cite{Thouless83,Zhou99,Brouwer98} 
in which charge distribution in a mesoscopic device is controlled by a slowly-varying quantum interference pattern. The same charge loading and unloading picture as shown in 
Figure~\ref{fig:ono} applies to quantum pumping of integer charge \cite{Kashcheyevs2004} where the discrete level spacing between quasi-bound resonant states
determines the addition energy. 
Under strict adiabaticity conditions (defined more accurately in Section~\ref{sec:modIntro}) an adiabatic pump requires at least two parameters to produce a non-zero dc current \cite{Brouwer98}. In the above example of parametric modulation of $V_{g1}$ and $V_{g2}$ this condition requires the pumping contour to  enclose a finite area in order to yield a finite pumped charge per period in the low-frequency limit. 
(Note that a turnstile, being subjected to a finite bias during operation, is a non-adiabatic device even without modulation.)
Hence any dc current produced by a single periodically varied parameter is a sign of essentially non-adiabatic operation \cite{Vavilov2001,moskalets2002B,
Torres2005,Arrachea2005erratum}. % , i.e.\ significant departure of the charge on the QD from instantaneous equilibrium. 

A clear example of single-gate pumping (although not yet in a quantized regime) has been demonstrated by Kouwenhoven \emph{et al.}\ \cite{kouwenhovenB91}, again with the same device design as presented in Figure~\ref{fig:Kouw}. The pumping scheme is shown schematically in Figure~\ref{fig:Kouw2}(b).
Only the gate voltage creating the barrier to the source is oscillating and no bias voltage is applied. During the first part of the cycle the source barrier increases and the conduction band bottom in the dot is  raised according to the capacitance between barrier gate and dot. As a result the electrochemical potential in the quantum dot $\mu_\mathrm{qd}$ is lifted above both $\mu_S$ and $\mu_D$, as shown by the hatched region. The electrons are raised in energy as the barrier to the source grows. During the second part of the cycle the source barrier is increased even further and electrons tunnel out of the dot with a preference to the drain reservoir. It is crucial that raising the energy of electrons (and hence switching the destination lead form source to drain) happens faster than  tunnelling out of the QD, otherwise the extra charge will be lost back to the source immediately once $\mu_\mathrm{qd}$ is raised above $\mu_S$. This non-adiabatic delay of tunnelling \cite{Kaestner2007c} is the key to efficient current generation by single-parameter modulation. Lowering the left barrier  during the third part of the cycle fills the dot again with electrons from the source  reservoir. Repeating the cycle results in pumping currents shown in Figure~\ref{fig:Kouw2}(c) for the device of Figure~\ref{fig:Kouw}. Peaks in the measured  current appear for an optimal setting of dc voltages on gates 1 and 2, i.e.\ when the fixed barrier is just in pinch-off and the other barrier oscillates around pinch-off. The direction of the pump current is reversed when switching the gate to which the rf signal is applied. The maximum of the current depends on the rf amplitude. Corresponding pump currents should be particularly robust against drain bias variations, as the current is determined only by the success of charge capturing from the source~\cite{fujiwara2008}.

\begin{figure}
\includegraphics[width=8cm]{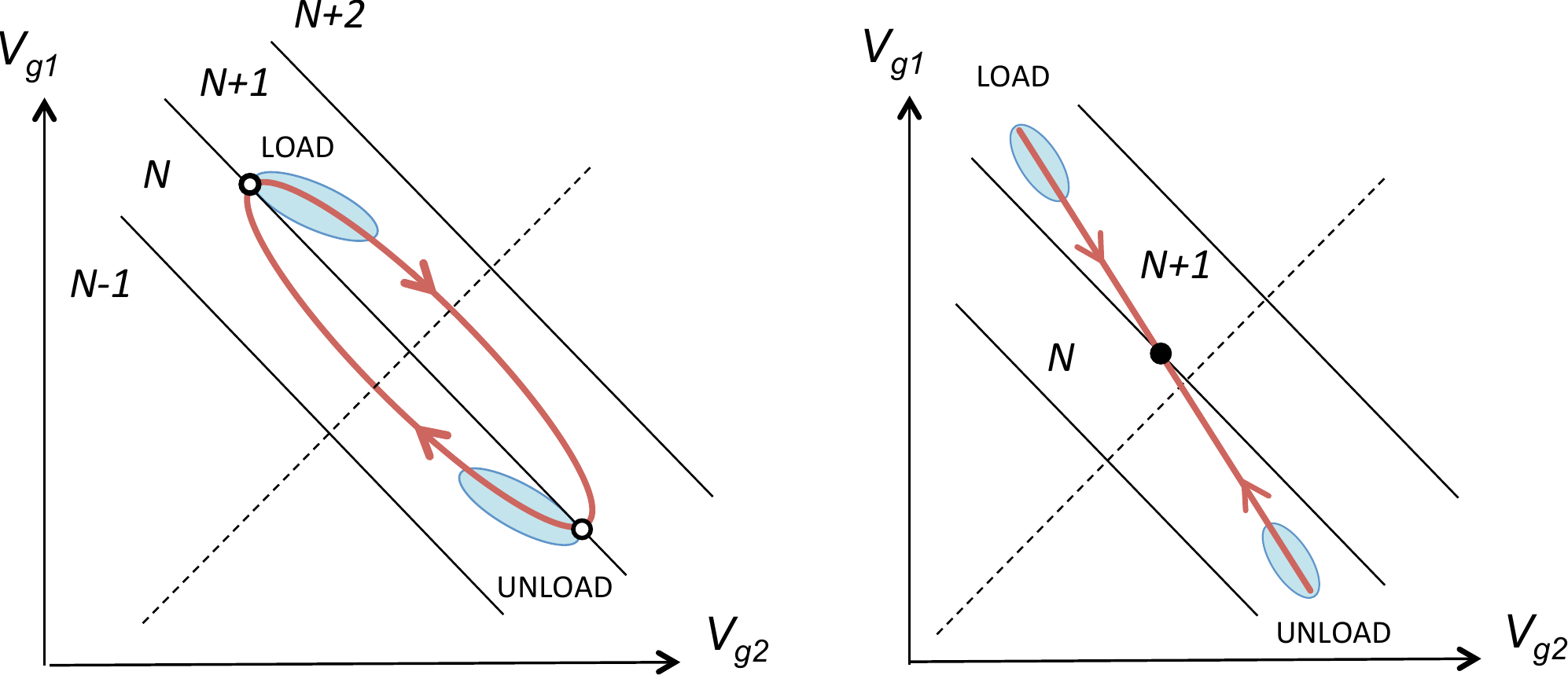}
\caption{Contrasting the adiabatic (a) and the single-parameter (b) pumping schemes for quantized charge transfer by entrance and exit gate modulation, controlled by voltages $V_{g1}$ and $V_{g2}$, respectively, through a tunable-barrier QD.}
\label{fig:ellipses}
\end{figure}
The essential features of the adiabatic versus the single-parameter pumping can be contrasted in a pair of schematic diagrams shown in Figure~\ref{fig:ellipses}. A series of anti-diagonal lines in each of the 2D plots
show the charge stability diagram separating regions of well-defined equilibrium number of electrons on the QD. Besides affecting the energy level of the QD, each gate strongly modulates the corresponding barrier with more positive voltage corresponding to a more open gate, similar to the design illustrated in Figure~\ref{fig:ono}. The elliptic contour in Figure~\ref{fig:ellipses}(a) is chosen for transfer of a single charge from the left (via barrier 1) to the right (via barrier 2) in the adiabatic limit; it will produce quantized current if the system is allowed to spend enough time 
in the regions of the intended charge loading and unloading (marked by small shaded ellipses in Figure~\ref{fig:ellipses}(a)).
The zero-area contour in Figure~\ref{fig:ellipses}(b)  defines a single-parameter pump (the actual parameter is a linear combination of $V_{g1}$ and $V_{g2}$ which depends on the ratio of the $\pi$-shifted modulation amplitudes). 
 The main difference between the two modulation schemes is the condition on the barrier transparency at the crossing points between the pumping contour and 
the boundary separating two charge configurations (marked by circles in Figure~\ref{fig:ellipses}): for adiabatic pumping (open circles in Figure~\ref{fig:ellipses}(a)) the QD at the crossing needs to be as \emph{open} as possible to allow proper charge equilibration with the appropriate contact while for the essentially non-adiabatic scheme (filled circle in Figure~\ref{fig:ellipses}(b)) the QD must be sufficiently \emph{closed} to induce non-adiabatic blockade and prevent unintended gain or loss of an electron as the topmost electron energy in the QD crosses the electrochemical potential of the contacts. These opposite design goals determine specific trade-offs in the choice of 
the modulation scheme for optimal quantized operation.  One advantage of the decoupling-oriented non-adiabatic scheme illustrated in Figure~\ref{fig:ellipses} is a greater freedom during the ``transit phase''of the cycle when the dot is effectively isolated and the dominating lead-dot coupling is being gradually switched for left to right (and vice versa while transiting in the opposite direction).
However, any scheme must allow sufficient time for charge loading and unloading and hence benefits from high-fidelity tuning of the barrier transparency.
We also note that once more than one modulation parameter is involved (e.g., the phase shift between two harmonic driving signals is not an integer multiple of $\pi$) there is no sharp boundary between the adiabatic and the non-adiabatic schemes, a continuous crossover is possible \cite{Croy2012a}.

\begin{figure}
\includegraphics[width=\smallfig]{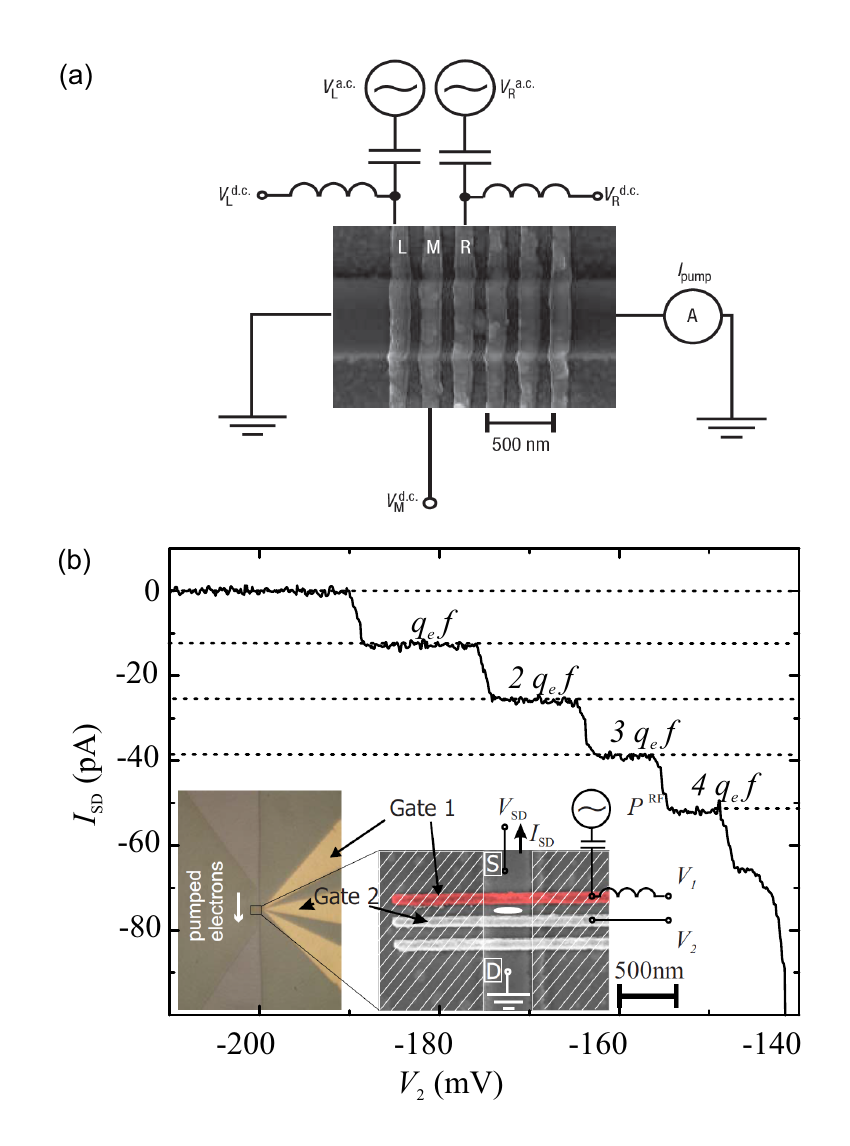}
\caption{(a) Scanning electron micrograph showing the GaAs-AlGaAs-based tunable-barrier device with etched wire and active metallic gates labeled R, M and L, as used by Blumenthal~\emph{et al.}~\cite{blumenthal2007a}.  (b) Single-gate quantized pumping, adapted from Kaestner~\emph{et al.}~\cite{kaestner2007b}. Quantized current generation by voltage modulation applied to gate 1 versus the dc voltage applied to gate 2 measured for the device shown in the insets.}
\label{fig:blumenthal}
\end{figure}

Single-parameter pumping over a barrier tuned into pinch-off has been demonstrated in the \emph{quantized} regime by Blumenthal~\emph{et al.}~\cite{blumenthal2007a}, using a device implemented in a GaAs-AlGaAs heterostructure. The three gates defining the dot are labeled L, M and R in Figure~\ref{fig:blumenthal}(a). They achieved quantized charge pumping by adding a small harmonic voltage modulating the drain barrier (R), shifted in phase by $180^\circ$ with respect to the large-amplitude modulation of the source barrier (L) ensuring that the drain barrier is kept in pinch-off during the whole cycle. Tuning parameter has been the middle gate (M), acting as plunger. The quantized current plateau is in agreement the exact value of one electron per cycle with an uncertainty of the order of $10^{-4}$ at 547$\,$MHz at a temperature of 300$\,$mK.

Realization of quantized electron  pumping  by single-\emph{gate} modulation has subsequently been demonstrated by Kaestner \emph{et al.} \cite{kaestner2007b, Kaestner2007c} and Fujiwara \emph{et al.} \cite{fujiwara2008}. The technological and conceptual advancement brought by the simplicity of this scheme has stimulated most of the recent developments which the subsequent sections of this Report are devoted to.

The GaAs-based devices used for the first demonstration of single-gate quantized current generation~\cite{kaestner2007b, Kaestner2007c} are shown in Figure~\ref{fig:blumenthal}(b). A sinusoidal modulation was added exclusively to gate 1. Tuning $V_1$ and $V_2$ applied to gate 1 and 2 respectively, a quantized current is generated (the pumping mechanism is discussed in great detail in Section~\ref{sec:currModel}). The figure shows the $I$-$V$ trace for 80$\,$MHz modulation, but the quantization persisted up to 800$\,$MHz. The measurement has been carried out at a temperature of 300$\,$mK. 
Single-gate GaAs-based devices have been further optimized for high speed and precision %\cite{giblin2012,Fletcher2012,seo2014} 
 using strategies discussed in Section~\ref{chap:metrology}.
%A discussion on optimization strategies can be found in Section~\ref{chap:metrology}. 
Giblin \emph{et al.} have traceably measured a single-gate operated pump at $I=150\,$pA which agreed with the quantized value within the measurement uncertainty of $1.2\,$parts per million (ppm)~\cite{giblin2012}.

Silicon-based devices used to realize an analogous single-gate pumping scheme~\cite{fujiwara2008} are similar to that shown in Figure~\ref{fig:BarrMod_Devs}(b). As discussed in Section~\ref{sec:BarrierMod} devices of this type can operate at a much higher temperature, in this case at 20$\,$K. Pulsed modulation of the source barrier has been used with a rise time $t_\mathrm{rise}$ and the duty cycle of the pulse signal being 2$\,$ns and 0.5, respectively. At frequency of $f =2.3\,$GHz and operating on the third plateau, a metrologically relevant (see Section~\ref{sec:metro}) quantized current of 1.1$\,$nA was measured.

Tunable-barrier pumps with individual donor atoms playing the role of the QD have also been demonstrated in silicon nanowires. Lansbergen~\emph{et al.}\ report pumping through a number of individual donors~\cite{Lansbergen2012}. Operated at a few MHz the device shows quantized pumping up to $6\,q_ef$ at a temperature of $T=36\,$K. In addition, the ionization energy was shown to be electrically tunable from $\approx 25$ to $54\,$meV.  Pumping through a single donor atom at a much higher rate of 1$\,$GHz was demonstrated by Tettamanzi~\emph{et al.}~\cite{tettamanzi2014} at $T=4.2\,$K. Yamahata~\emph{et al.} argue in \cite{Yamahata2014} that the use of charge trap levels as the quantization-defining localized states may lead to higher operation frequencies and precision. By electrically controlling the capture and emission rates to and from a trap level the authors of \cite{Yamahata2014} have demonstrated quantized pumping up to the frequency of 3.5$\,$GHz with a transfer accuracy of about $10^{-3}$, limited by their measurement uncertainty. The device operated at a temperature of $T=17\,$K.

Devices using aluminium gates to accumulate electrons at a Si-SiO$_2$ interface have been used by Rossi \emph{et al.}~\cite{Rossi2014}. This structure is particularly suitable for enhanced tuning of  the electrostatic confinement on the QD. By exploiting this flexibility the authors demonstrate current reversal using a two-signal drive and pumping at 500$\,$MHz with an uncertainty below $50\,$ppm operating at a base temperature of $80\,$mK.

%%%%%%%%%%%%%%%%%%%%%%%%%%%%%%%%%%%%%%%%%%%%%%%%%%%%%%%%%%%%%%%%%%%%%%%%%%%%%%%%%%%%%%%%%%%%%%%%
\subsection{Tunable Josephson junction devices}
\label{sec:Joseph}

\begin{figure}
\includegraphics[width=\smallfig]{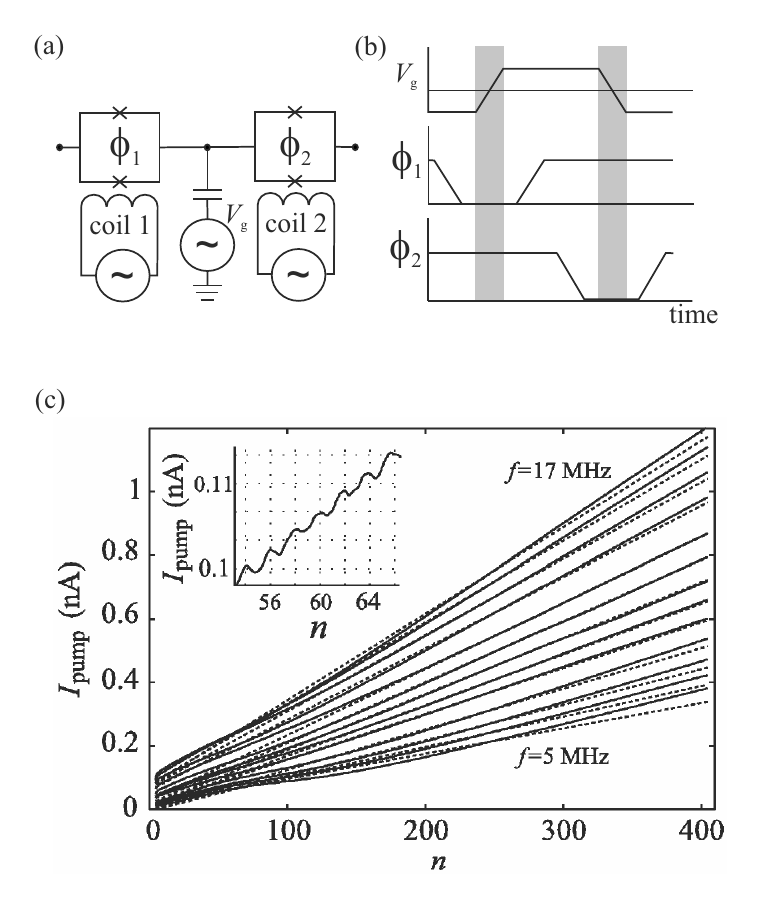}
\caption{(a) Schematic of the Cooper pair pump. Modulation parameters are the gate voltage $V_g$ and the magnetic fluxed $\phi_i$ of the SQUIDs, controlled by on-chip coils. (b) Diagram showing the synchronization of the modulation parameters. (c) Pumped current as function of frequency and gate modulation amplitude where $n$ refers to the ideal number of elementary charges transfered per cycle. The dotted theoretical currents $I = n e f + I_\mathrm{leak}$ were forced to match the experimental current at $n = 250$, with $I_\mathrm{leak}$ as a fitting parameter. The inset shows the expected periodicity of the current with gate modulation amplitude. Reproduced with permission from \cite{vartiainen2007}.
Copyright 2007, AIP Publishing LLC.}
\label{fig:SLUICE}
\end{figure}

In analogy to a QD connected to the leads by tunable semiconducting barriers, a superconducting island may be connected by two SQUIDs to the leads. The SQUIDs can be considered here as tunable Josephson junctions, i.e.,  as \emph{valves} that can be opened or closed for tunneling. The coupling is controlled by local magnetic fluxes $\phi_i$ to each lead $i$ using on-chip superconducting coils, which change the critical current of each of the SQUIDs as shown in Figure~\ref{fig:SLUICE}(a). Each pump cycle results in  transfer of Cooper pairs yielding twice the current compared with single electron-pumps operated at the same frequency. A corresponding scheme has first been proposed and experimentally demonstrated by Niskanen \emph{et al.} \cite{niskanen2003, niskanen2005}. The principle of operation is shown in Figure~\ref{fig:SLUICE}(b). Throughout the cycle at least one SQUID is closed (minimum critical current). The gate voltage is tuned to move Cooper pairs through the open SQUID (maximum critical current), shown as gray shaded regions in the Figure. Each operation cycle can transfer up to several hundreds of Cooper pairs, as demonstrated by Vartiainen \emph{et al.}~\cite{vartiainen2007}, leading to currents in the nanoampere range at pumping frequency of $10\,$MHz, see Figure~\ref{fig:SLUICE}(c). The pumped current as function of frequency and amplitude of the gate voltage modulation is consistent with quantized Cooper pair pumping. However, the voltage bias leads to leakage currents. Strategies to improve the accuracy have been explored by M\"ott\"onen \emph{et al.} \cite{mottonen2008} and Gasparinetti \emph{et al.} \cite{gasparinetti2012}.

%%%%%%%%%%%%%%%%%%%%%%%%%%%%%%%%%%%%%%%%%%%%%%%%%%%%%%%%%%%%%%%%%%%%%%%%%%%%%%%%%%%%%%%%%%%%%%%%
\subsection{Modulation by surface acoustic waves}
\label{sec:SAW}

\begin{figure}
\includegraphics[width=\smallfig]{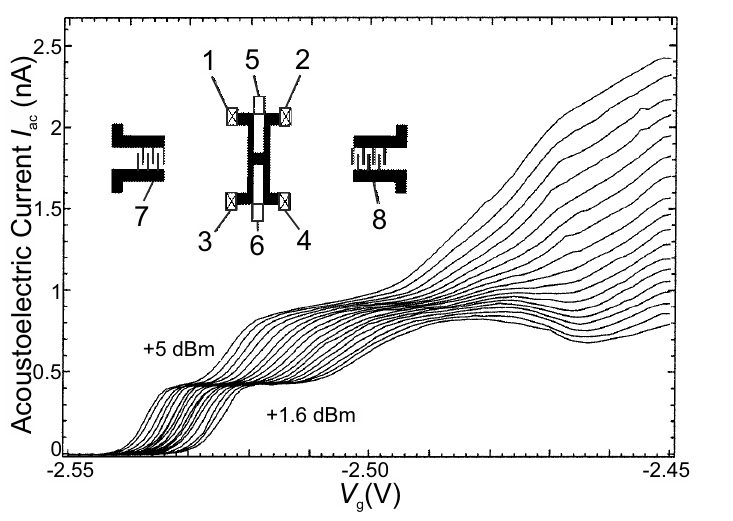}
\caption{Quantized current generated by SAWs interacting with a quasi-one dimensional channel for varying SAW power~\cite{Shilton1996}. Inset: schematic of the device; ohmic contacts are labeled by 1-4, split gates by 5 and 6, and SAW transducers by 7 and 8. The sample was implemented in a GaAs-AlGaAs heterostructure. The two transducers are separated by $4\,$mm. Copyright IOP Publishing. Reproduced by permission of the authors and IOP Publishing. All rights reserved.}
\label{fig:SAW}
\end{figure}

Owing to the piezoelectric properties of GaAs-based substrates surface acoustic waves (SAWs) on these substrates are accompanied by waves of electrostatic potential. The modulation of QD potentials by SAWs is a subject of ongoing experimental \cite{Shilton1996, Talyanskii1997, Cunningham1999, Cunningham2000, Ebbecke2000, Janssen2001, Robinson2002, Fletcher2003, ebbecke3, stotz2005, Ebbecke2005, kataoka2006a, kataoka2006b, ahlers2006, naber2006, astley2007, kataoka2007, schneble2007, Wurstle2007, Buitelaar2008, McNeil2011, Hermelin2011, Hermelin2013, Chen2013, he2014} and theoretical \cite{Aizin1998, Gumbs1999, Flensberg1999, maksym1, Galperin2001, Robinson2001, Aharony2002, Kashcheyevs2004, Buitelaar2008} research. 
Shilton \emph{et al.}~\cite{Shilton1996} have first shown quantized charge transport using this principle. The typical experimental arrangement is shown schematically in the inset of Figure~\ref{fig:SAW}: a quasi-one dimensional channel is defined by a split gate (labeled as 5 and 6 in the figure) in a GaAs-AlGaAs heterostructure and a SAW is launched in the longitudinal direction at frequency $f_\mathrm{SAW}$ from a transducer (labeled 7 and 8). The acoustoelectric dc current $I$ is measured between the contacts 1/3 and 2/4. Under appropriate conditions, $I$ exhibits a staircase plateau-like structure as function of the gate voltage (which controls the depletion of the channel) and of the SAW power, as shown in Figure~\ref{fig:SAW}. At the plateaus, the current saturates at quantized values $I = n q_e f_\mathrm{SAW}$, corresponding to the transfer of an integer number $n$ of electrons per each period of the SAW. The influence of factors such as source-drain bias~\cite{Shilton1996, Talyanskii1997,Cunningham2000,Gloos2004}, temperature~\cite{Shilton1996, Janssen2001, Fletcher2002}, power \cite{Shilton1996, Ebbecke2000, Gloos2004, ahlers2006}, perpendicular magnetic field~\cite{Shilton1996, Cunningham2000,he2014} and a weak counter propagating SAW beam~\cite{Cunningham1999,he2014} on the staircase structure and plateau quality have been studied experimentally.
%, with some qualitative tendencies match by simple one-dimensional tight-binding models \cite{Aharony2002,Kashcheyevs2004}.

The transducers typically consist of metallic finger pairs such that the SAW wavelength $\lambda$ is determined by the period of the transducer fingers. The most efficient generation of SAWs occurs when modulating the finger pairs at the resonance frequency $f_0 = v_\mathrm{SAW}/\lambda$. Here $v_\mathrm{SAW}$ is the velocity of the SAW which depends on the substrate, temperature, and other parameters. The resonance frequency for the GaAs-AlGaAs based device used by Shilton \emph{et al.}~\cite{Shilton1996} at a set base temperature of 300$\,$mK was $f_0 = 2728.6\,$MHz, as determined from the transmittance between transducers 7 and 8. Hence choosing $\lambda \approx 1\,\mu$m, i.e. a typical dimension of a QD in a GaAs-AlGaAs 2DEG, the corresponding modulation is much faster than most realizations discussed in Secs.~\ref{sec:plungerBarrier} and \ref{sec:BarrierMod}. The highest frequency at which quantized pumping could be observed was 4.7$\,$GHz by Ebbecke~\emph{et al.}~\cite{Ebbecke2000} also using the GaAs-AlGaAs heterostructure. Most experimental realizations indeed employ GaAs-AlGaAs because it combines the properties of piezoelectricity and low-dimensional electron transport. However, also insulating substrates such as LiNbO$_3$ or quartz have been considered, because of favourable piezoelectric or thermal properties. The conducting channel has been provided by carbon nanotubes~\cite{Ebbecke2004a, leek2006a, buitelaar2006, Wurstle2007}. In such a system, quantized pumping was first demonstrated by Buitelaar~\emph{et al.} \cite{buitelaar2006}, at a pump frequency of $f_\mathrm{SAW} \approx 3\,$GHz.

Most models \cite{Aizin1998, Gumbs1999, Flensberg1999, Robinson2001} for SAW pumps treat the electrons \emph{already localized} in a moving potential well (dynamic QD) and those belonging to the Fermi sea separately. The current is then determined by the loss of electrons from the dynamic QD at the stage of its formation \cite{Flensberg1999} and/or its subsequent motion \cite{Aizin1998, Gumbs1999, Flensberg1999, Robinson2001}. Quantization error mechanisms within these models (gradual backtunneling \cite{Aizin1998, Gumbs1999}, non-adiabaticity at formation stage \cite{Flensberg1999}, non-equilibrium classical dynamics \cite{Robinson2001}) consider energies that can significantly exceed the Fermi energy in the remote reservoirs, putting the SAW pumps in the class of strongly non-adiabatic pumps (see discussion in Section~\ref{sec:modIntro} below). Quantum dynamics in the non-equilibrium \cite{maksym1} and adiabatic limits \cite{Aharony2002,Kashcheyevs2004} has been analyzed within single-particle models of 1D time-dependent potentials. The latter calculations do not require the presence of a dynamic QD at all times and confirm that localized electronic states are responsible for the quantized transport.  Experimentally, driving conditions for both non-adiabatic \cite{kataoka2006b, crook2010, he2014} and close-to-equilibrium \cite{ahlers2006} regimes have been identified.

In all above realizations the fixed link between the QD size and the modulation frequency as well as the limitation to harmonic modulation have been seen as a challenge for device optimization~\cite{blumenthal2007a}, or for improving the relatively low yield \cite{Fletcher2003, ebbecke3}. Experiments dealing with the latter issue have combined gate defined QDs with SAW modulation~\cite{ebbecke3}. Another challenge represents the requirement for a large driving power: in order to obtain flat current plateaus large microwave powers in excess of 10$\,$dBm have to be applied to the SAW transducer. Experimental studies by Janssen and Hartland~\cite{Janssen2000a, Janssen2001} indicate that rf heating may be the limiting factor in improving the accuracy in SAW based pumps and that the real electron temperature for their investigated device lies around 12.5$\,$K keeping the insert temperature at 1.5$\,$K. In this experiment the pumping current at the middle of the current plateau was determined to be $(70 \pm 14)\,$fA below the quantizted value of 430.61$\,$pA~\cite{Janssen2000a}. To date this represents the most precise current value for SAW-pumps. The measurement by Utko~\emph{et al.} \cite{Utko2006} allowed to resolve the power deposited by the SAW itself. As the dominant contribution of power deposition they identify overall rf heating rather than losses due to SAW inside the device, which would be far more difficult to remove.

%\input{experiments}
%%%%%%%%%%%%%%%%%%%%%%%%%%%%%%%%%%%%%%%%%%%%%%%%%%%%%%%%%%%%%%%%
%%%%%%%%%%%%%%%%%%%%%%%%%%%%%%%%%%%%%%%%%%%%%%%%%%%%%%%%%%%%%%%%
%%%%%%%%%%%%%%%%%%%%%%%%%%%%%%%%%%%%%%%%%%%%%%%%%%%%%%%%%%%%%%%%

\section{Elements of quantitative modelling\label{chap:quantModel}}
The discussion in the previous section has been largely qualitative owing to the diversity of technological and physical factors affecting the operational envelope of tunable-barrier single-electron current sources. Recent progress in the field has centred around a particular pumping scheme that singles out the  non-adiabatic charge capture as the crucial phase of the pumping cycle in a tunable-barrier QD. This approach reveals a certain degree of simplicity and universality which we aim to explain in Sections ~\ref{sec:modIntro} and \ref{sec:currModel} below, from the theoretical and the experimental viewpoints, respectively.
We then use the non-adiabatic capture statistics and related fitting formulas to review the experimental lessons  learnt recently from the average current  (Section~\ref{sec:currModel}), shot noise (Section~\ref{sec:noise}), and
electron counting (Section~\ref{sec:CountStatModel}).

\subsection{Theory background}
\label{sec:modIntro}

Essential aspects of quantized charge pumping can be understood within the framework of rate equations for a single non-equilibrium degree of freedom -- the number $n$ of charges confined on the quantum dot. In a Markov approximation and ignoring the quantum broadening effects, the kinetic equation for 
probability $P_n(t)$ to find $n$ charges at time $t$ can be written as
\begin{eqnarray}  
  \frac{d}{dt} P_n & =   -   \Gamma_{n} \left   [ \bigl ( 1-f(\mu_n) \bigr ) P_n -f(\mu_n)  P_{n-1} \right ]   \nonumber \\
    + &  \Gamma_{n+1}   \left [\bigl ( 1- f(\mu_{n+1}) \bigr ) P_{n+1}- f(\mu_{n+1}) P_n \right ] . \label{eq:det:kin1}
\end{eqnarray}
Here $f(E) = 1/\{ 1+\exp [(E-\mu)/kT] \}$ is the Fermi function, $\mu$ is the electrochemical potential of the lead, and $\Gamma_n = W_{n-1}^{+} + W_{n}^{-}$ is the sum of electron addition ($+$) and removal ($-$) rates for charge fluctuation between $n-1$ and $n$ confined electrons. 
We shall apply \eqref{eq:det:kin1} to parts of the pumping cycle where coupling to only one of the leads is relevant, hence the lead index ($S$ for the source, $D$ for the drain) is omitted in this section.

The addition and removal rates are connected by the detailed balance condition 
\begin{equation} \label{eq:detailedW} 
W_{n}^{-}/W_{n-1}^{+}= e^{(\mu_n-\mu)/kT}  ,
\end{equation}
which defines
the electrochemical potential $\mu_n$ of the dot with $n$ electrons.
The addition energy $\mu_n-\mu_{n-1}$ is typically dominated by the capacitative charging energy $E_c=e^2/C$ for $n \gg 1$ but becomes enhanced and dependent
on the shape of the quantum dot for the last few electrons~\cite{kouwenhoven2001}. 
It is important to stress that identifying $T$ in \eqref{eq:detailedW} with thermodynamic temperature is justified only if thermal equilibrium is established fast enough, over time scales shorter than $(\Gamma_n)^{-1}$ (fast thermalization limit \cite{LiuNiu1997}).

For tunneling-dominated transport through the barrier, the rates can be calculated by applying the Fermi Golden rule to the tunnelling Hamiltonian,
\begin{eqnarray} 
W_n^{-} = & & \frac{4 \pi^2}{h} \int \rho_{\rm QD} (E) \rho_{L} (E) |\mathcal{V}_T(E)|^2  \nonumber \\
  & & \times f(E+\mu-\mu_n) \left [ 1 - f(E) \right ] d E \label{eq:microscopic}
\end{eqnarray}
with the addition rate $W_{n-1}^{+}$ given by the same equation \eqref{eq:microscopic} but with  $f$ replaced by $1-f$.
Here $\rho_{\rm QD}(E)$ and $\rho_{L} (E)$ are the densities of state in the quantum dot (including spin degeneracy) and the lead,
respectively, and $\mathcal{V}_T(E)$ is the tunnelling matrix element, all averaged over mesoscopic fluctuations at the single-electron energy $E$
(justified for small level spacing, $\rho_{\rm QD} kT \gg 1$ and fast thermalization on the QD).

An important energy scale for tunable tunnel barriers, henceforth denoted $\Delta_b$ and known as ``transverse energy'' \cite{Pekola2013}, 
characterizes the rate of exponential growth of transmission probability with energy, 
$\rho_{\rm QD} (E) \rho_{L} (E) |V_T(E)|^2 \propto e^{E/\Delta_b}$. 
Convergence of the integral in the expression for the tunnelling rate \eqref{eq:microscopic} requires $kT < \Delta_b$; 
at higher temperatures  
tunneling crosses over to thermal hopping,  and the charge fluctuation rates at $kT > \Delta_b$ are determined by 
activation above the classical barrier height $E_b$, i.e. $W_{n}^{-} \propto e^{-(E_b-\mu_n)/kT}$ and
$W_{n-1}^{+} \propto e^{-(E_b-\mu)/kT}$ (see Eqs.~(5) and (6) in \cite{Matveev1996}). 
For tunnelling, a single-electron WKB approximation for one-dimensional rectangular barrier leads to an estimate of $\Delta_b$ as \cite{Zimmerman2004,Fletcher2012}
$\Delta_b = h\sqrt{(E_b-\mu_n)/(2 m^{\ast})} /(2 \pi L)$, where $L$ is the barrier length and $m^{\ast}$ is the effective electron mass, whereas
for a parabolic barrier model \cite{Matveev1996,YamahataPRB2014} the transverse energy $\Delta_b$ is independent of the barrier height $|E_b-\mu_n|$ \cite{Kemble1935}.

A change $\Delta V_g$ in the gate voltage $V_g$ controlling the tunnel barrier has a two-fold effect on $\Gamma_n$: 
a \emph{plunger function} consisting of shifting
the energies on the dot, $\mu_n \to \mu_n + q_e \Delta V_g C_{\rm{g-QD}}/C$ (here $C_{\rm{g-QD}}$ 
is the capacitance between the gate and the QD and $C$ is the total capacitance of the QD), and a \emph{barrier function}, affecting the tunnelling matrix element.
The barrier function can be approximated analytically as $|\mathcal{V}_T(E)|^2 \to |\mathcal{V}_T(E)|^2 \exp (- \Delta V_g  d E_b/d V_g /\Delta_b) $ where $d E_b/d (q_e V_g) $ is the lever arm factor for the gate voltage affecting the top of the potential barrier.
%\sout{Note that in the low-temperature limit, the factor $d  E_b/d V_g$  is related  to the non-thermionic subthreshold slope $S$ of the barrier acting as an FET, $d  E_b/d V_g = \Delta_b \ln 10/ S$}. 
Additional complexity to energy and voltage dependence of the charge exchange rates through tunable semiconductor barriers may come from mesoscopic transport paths beyond direct tunneling. Such paths may involve hopping or resonant conductance via individual donors \cite{Koenraad2011}, interface charge traps~\cite{Ebbecke2005,YamahataPRB2014} or other disorder-induced localized states (e.g., resonances in gated graphene~\cite{guettinger2011}).

In the  tunneling limit, $k T  \ll \Delta_b$, 
the integral \eqref{eq:microscopic}  can be approximated 
\begin{equation} \label{eq:orthodox}
  W_n^{-} = (G_T/e^2) (\mu_n-\mu)/(1-e^{-(\mu_n-\mu)/kT}) \, ,
\end{equation} 
where $G_T$ is the tunnelling conductance of the barrier, 
$G_T \approx (4 \pi e^2/h)% \overline{
\rho_{\rm QD} \rho_{L}  |\mathcal{V}_T|^2
%}
$, averaged over a bias window 
$k T < |e V_{\rm bias}| < \Delta_b$. Equations \eqref{eq:det:kin1} with rates \eqref{eq:orthodox}  are the basis for the sequential-tunnelling (``orthodox'') theory of Coloumb blockade~\cite{grabert1PBI}, widely used for  simulation of fixed-barrier devices with metallic islands~\cite{Flensberg1999a,Pekola2013} and, more recently, 
silicon-based pumps with exponentially tunable $G_T$ \cite{Chan2011a, Ray2014}. 

For low temperatures and strong confinement, quantum effects beyond tunneling are expected to play an increasingly important role. When level spacing exceeds the thermal energy scale, mesoscopic effects make addition energies and escape rates sensitive to specific wave functions of the $n$-body states on the QD \cite{Aleiner2002}.  For the ground state of the last one or two electrons in a predictable potential,  numerical solution of the Schr{\"o}dinger equations  \cite{Aizin1998,Gumbs1999} or numerical lattice methods \cite{Sim1997,Kaestner2007c,Fletcher2012,seo2014} can yield useful information on parametric dependence of the tunable tunneling rates.

Modeling of parametrically-driven transport relies on time-scale separation \cite{Jauho1994,Esposito2012}: changes in gate voltage are assumed to affect $\Gamma_n(t)$ and $\mu_n(t)$ instantaneously (i.e., these are \emph{fast} variables) while $n$ can be either a slow or a fast variable, depending on relation of the equilibration rates $\Gamma_n(t)$
to the external modulation speed.  
In \emph{adiabatic} pumping, $P_n(t)$ stays close to the rate-independent quasi-static equilibrium distribution $P_n^{\rm{eq}}(t) \propto \exp \sum_{m=1}^n \{-[ \mu_n(t)-\mu ]/kT\}$ during the whole cycle ($P_n^{\rm{eq}}$ is  the solution to \eqref{eq:det:kin1} with the l.h.s.\ set to zero).
Typical realizations of adiabatic pumps with two barriers tuned out of phase can be found in~\cite{Ono2003a,jehl2013} (see Section~\ref{sec:plungerBarrier}).
In contrast, the \emph{non-adiabatic} quantized charge pumping scheme \cite{Kaestner2007c} aims to decouple the quantum dot from the source
before coupling it to the drain, thus permitting the number of electrons on the dot to differ substantially from the equilibrium value formally expected from the instantaneous position in the charge stability diagram. 

A general strategy for simulating a charge pump  consists of obtaining 
a periodic solution of the appropriate kinetic equation (equation \eqref{eq:det:kin1} being one of the simplest examples) along a particular closed contour in the
parameter space and computing the corresponding period-averaged sequential tunnelling current \cite{Aleiner1998,Brouwer98,Zhou99,Shutenko00,Makhlin01,moskalets2002B,Kashcheyevs2004,Aono2004,Splettstoesser05,Sela05,Kaestner2007c,Fioretto08,Arrachea08, Leicht2009,Cavaliere09,battista2011,Chan2011a,Croy2012a}. 
Universal, analytic results  are possible only in special limits, 
of which a particularly useful one is the statistics of charge capture \cite{liu1993,Flensberg1999,Robinson2001,Zimmerman2004,kaestner2010a,Fricke2013,YamahataPRB2014} (see also an analytic solution for time-limited emission with constant 
rates in \cite{astley2007,miyamoto2008,Yamahata2014,Kashcheyevs2014}). To model charge capture in a QD by a closing tunable barrier, we follow \cite{Fricke2013} and consider 
a close-enough-to-equilibrium initial state of the dot  at $t=t_0$, when it is well connected to the source lead. At $t>t_0$ 
a linear ramp of a gate voltage leads to exponential reduction of $\Gamma_n(t) = \Gamma_n(t_0) e^{-(t-t_0)/\tau}$ up to $t=t^{\ast}$ when
the coupling to all leads is small enough to be negligible. The characteristic decoupling time $\tau$ is set by the ramp rate, and the barrier function
of the gate. It is assumed that $\Gamma_n(t_0) \gg \tau^{-1}$ so that initially
the evolution of $P_n(t)$ follows closely the instantaneous adiabatic values $P_n^{\rm{eq}}(t)$.

Due to barrier-plunger crosstalk, the electrochemical potentials on the dot drift during closing at a certain rate $d \mu_n(t)/dt$.
Whether this drift is important enough to dominate the capture error mechanism, depends on the value of the plunger-to-barrier ratio, 
$\Delta_{\rm ptb} = \tau | d \mu_n(t)/dt |$, as first discussed by Kashcheyevs and Timoshenko in~\cite{Kashcheyevs2012a}.
If $\Delta_{\rm ptb} \ll kT$ then the Fermi functions in \eqref{eq:det:kin1} do not change appreciably during the decoupling process, and a sudden approximation is appropriate. Detailed analysis of this limit \cite{Fricke2013} in case of well-defined quantization, $\mu_{n+1}-\mu_{n} \gg kT$,  leads to
 the following generalized grand canonical distribution for $P_n(t^{\ast})$:
\begin{eqnarray} 
   P_n & = \left[1- f\bm{\left(}\tilde{\mu}_{n\!+\!1}^{}\bm{\right)}\right] \prod_{m=1}^{n} f\left(\tilde{\mu}_m\right) \nonumber \\
  &  \approx  f\left(\tilde{\mu}_{n}\right)  - f\left(\tilde{\mu}_{n+1}\right)  .\label{eq:gradtofit}
\end{eqnarray}
Here $\tilde{\mu}_n$ is the electrochemical potential $\mu_n(t^c_n)$ of the $n$-th charge state frozen at a  sufficiently well-defined decoupling moment
$t^c_n$ such that for $t>t^c_n$ both rates affecting $P_n(t)$ in \eqref{eq:det:kin1}, $\Gamma_{n}$ and $\Gamma_{n-1}$, drop below $\tau^{-1}$, 
thus effectively disengaging the $n$-th charge state from the detailed balance. 
(Technically, $t^c_n$ can be defined by $\int_{t^c_n}^{t^{\ast}}\!\Gamma_n(t)\,dt\!=\!1$~\cite{Kashcheyevs2012a,Fricke2013}.)
Assuming the drift of the energy levels and the rate of reduction of the matrix elements to be the same for subsequent charge states (ie., $n$-independent $\tau$ and $\Delta_{\rm ptb}$),
the rate $\Gamma_{n+1}(t)$ reaches the value of $\Gamma_n(t^c_n)$  at a later time $t=t^c_{n+1} = t^c_{n} +\tau \ln [\Gamma_{n+1}(t)/\Gamma_{n}(t) ]$. The shift of $\mu_{n+1}(t)$ during the time from $t^c_{n+1}$ to $t^c_{n}$ increases the effective energy gap by $(\tilde{\mu}_{n+1}-\tilde{\mu}_{n})-(\mu_{n+1}-\mu_{n})=\Delta_{\rm{ptb}} 
 \ln (\Gamma_{n+1}/\Gamma_{n})$. In the extreme sudden decoupling limit the latter difference is negligible compared to the thermal broadening, $\Delta_{\rm{ptb}} \ln (\Gamma_{n+1}/\Gamma_{n}) \ll kT $, and \eqref{eq:gradtofit} reduces to the grand canonical distribution $P^{\rm eq}_n(t^c)$ corresponding to thermodynamic equilibrium with parameter values fixed at time $t=t^c \approx t^c_{n} \approx t^c_{n+1}$ (still assuming, however, that $T$ is an adequate measure of local temperature).

In the opposite limit of large plunger-to-barrier ratio, the non-equilibrium dynamics during the decoupling process is essential and the final 
probability distribution differs strongly from the thermal limit \eqref{eq:gradtofit}. For $\Delta_{\rm ptb} \gg kT$ and $d \mu_n/dt >0$,
the dominating process is the loss of electrons into the empty states in the lead once $\mu_n(t)$ exceeds the sufficiently sharp electrochemical potential of the source $\mu$, see \eqref{eq:detailedW}. 
This regime is known as the decay cascade limit \cite{kaestner2010a,Fricke2013,Kashcheyevs2014}  and results in the following
probability distribution for the captured charge: 
\begin{eqnarray} 
P_n & = e^{ -X_{n}} \prod_{j=n+1}^{\infty} \left( 1 - e^{-X_{j}} \right) 
\nonumber \\ 
& \approx e^{-X_n}-e^{-X_{n+1}} . \label{eq:deacyfull}
\end{eqnarray}
Here $X_n =\int_{t_0}^{t^{\ast}} W_n^{-} dt$ is the electron escape rate integrated over the part of the pumping cycle corresponding to the
gradual decoupling from the source lead concurrent with lifting of the quantum dot above the Fermi sea. Since for $\Delta_{\rm ptb} \gg kT$
the Fermi functions are sufficiently sharp, and $W_n^{-}(t)=\{1-f[\mu_n(t)]
\} \Gamma_n(t)$, the escape rate integral can also be evaluated as
$X_n =\int_{t_{n}^b}^{t^{\ast}} \Gamma_n(t) \, dt$ where $t_{n}^b$ is the backtunneling onset time, $\mu_n (t_{n}^b )= \mu$.
For a linear $\mu_n(t)$ and exponential $\Gamma_n(t)$ the parameters of \eqref{eq:deacyfull} and \eqref{eq:gradtofit} are
connected as $X_n = \exp [ (\tilde{\mu}_n-\mu)/\Delta_{\rm ptb} ]$. Empty QD is formally assigned $X_0=0$ and $\tilde{\mu}_{0} =-\infty$.

The decay cascade distribution \eqref{eq:deacyfull}, derived under the condition $X_n \ll X_{n+1}$, is peaked at $n_0$ if $X_{n_0} \ll 1$ (the non-adiabatic loss of electrons, once
the  state $n_0$ is out of equilibrium, is negligible) and simultaneously $X_{n_0+1} \gg 1$ 
(the escape rate has been sufficient to get rid of the unwanted $(n_0+1)$th electron).
A dimensionless ratio characterizing the sharpness of the distribution \cite{kaestner2010a}, $\delta_n = \ln (X_n/X_{n-1})$,
has contributions both from the disparity of instantaneous escape rates and from 
the delay of the onset of backtunneling due to finite charging energy \cite{VKJT2014}, 
$\delta_n = \ln (\Gamma_n/\Gamma_{n-1}) + (\mu_n-\mu_{n-1})/\Delta_{\rm ptb}$. In the limit of plunger-to-barrier ratio exceeding
the charging energy, $\Delta_{\rm ptb} \gg \mu_{n}-\mu_{n-1}$, as originally considered in \cite{kaestner2010a}, the first term
in the above expression for $\delta_n$ dominates.
Note that the decay cascade distribution \eqref{eq:deacyfull} is not limited to tunneling barriers, but applies to thermally-activated
escape as well \cite{kaestner2010a}, as long as $kT$ is much smaller than the corresponding $\Delta_{\rm ptb}$.

Conditions for the crossover between the equilibrium and the decay cascade distributions for charge trapping in Si tunable-barrier quantum dots 
have been recently analyzed theoretically by Yamahata \emph{el al.} \cite{YamahataPRB2014,Y2014err}.
Under assumptions similar to the ones underlying \eqref{eq:microscopic} (in particular, negligible mesoscopic effects), they have expressed the plunger-to-barrier ratio for  tunneling, 
$\Delta_{\rm ptb}=g \, \Delta_b$,
and for thermally-activated hopping, $\Delta_{\rm ptb}=g \, k T$, in terms of a single combination of capacitative coupling factors 
which in our notation reads $g =( d \mu_n/d V_g)/(d E_b/d V_g -d \mu_n/d V_g)= d \mu_n/d (E_b-\mu_n)$. Within the same model one can relate the escape rate 
ratio to the addition energies, $\ln(\Gamma_n/\Gamma_{n-1})= (\mu_n-\mu_{n-1})/\Delta_b$ (tunneling) or $(\mu_n-\mu_{n-1})/kT$  (hopping), and 
thus estimate $\delta_n = (g+1) (\mu_n-\mu_{n-1}) / (g \Delta_b) $ for tunneling ($k T < \Delta_b$)  and $\delta_n = (g+1) (\mu_n-\mu_{n-1}) / (g kT) \approx (\mu_n-\mu_{n-1})/k T $ for  
hopping ($k T > \Delta_b$) cascades.

The presented analysis of the sudden \eqref{eq:gradtofit} or gradual \eqref{eq:deacyfull} breakdown  of detailed balance for the single non-adiabatic variable $n$ has relied on the general kinetic equation \eqref{eq:det:kin1} and the time-dependence of the rates $W^{\pm}_n(t)$. If the rates themselves are adiabatic, i.e.\ respond quasi-statically to changes in the electrostatic potential driven by external pumping parameters, then it is not difficult to predict the changes to the capture statistics due to varying decoupling speed. For example, increasing the gate voltage modulation rate by a factor of $\lambda$ would change $\tau \to \tau/\lambda$, $X_n \to X_n/\lambda$ and keep $\Delta_{\rm ptb}$ unchanged (an example \cite{fujiwara2008} of such scaling is discussed in Section~\ref{sec:currModel}).

The idealization of time-scale separation underlying the concepts of adiabatically modulated rates and a well-defined  temperature are often hard to verify experimentally, especially for strong and fast modulation.
Several mechanisms for non-adiabatic excitation of additional degrees of freedom beyond the electron number that are relevant for tunable-barrier pumping have been discussed in the literature. On a single-electron level, these excitations may be driven by (a) loading of hot electrons into the excited states \cite{LiuNiu1997,fricke2011} of an empty QD as coupling to a source lead with a mismatched electrochemical potential is enabled; (b) wave-functions of the electrons not having enough time to adapt to the changing shape of the confining potential \cite{kataoka2011,Fletcher2012} or the growing height of the tunneling barrier \cite{liu1993,Flensberg1999,Kashcheyevs2012a}.  In particular, non-adiabatic excitation of electrons in the leads 
due to an exponentially decreasing tunnel matrix element $\mathcal{V}_T(t)$ has been linked to the break down of the Markov approximation 
underlying  \eqref{eq:det:kin1}. The corresponding energy scale for dynamic quantum broadening $h/\tau$ is expected to compete with $kT$ and $\Delta_{\rm ptb}$ \cite{Kashcheyevs2012a}.
On the level of a decay cascade from an initial many-electron state on the QD, recoil energy of the electron(s) remaining on the dot after the last escape event was suggested as the precision-limiting factor for capture statistics $P_n$, based on classical dynamics simulations of a SAW-created dynamic QD \cite{Robinson2001}. Tentative agreement of the latter results to the decay cascade distribution \cite{kaestner2010a} suggests that non-adiabatic excitations inside the QD may still be accommodated in the  Markovian framework \cite{%Cavaliere09,
Esposito2012} by replacing $\mu_n$ and $kT$ in the detailed balance condition \eqref{eq:detailedW} with appropriate effective values.
Nevertheless, quantitative modeling of non-adiabatic effects in charge capture remains among important open issues for theory.

\subsection{Pumping currents}
\label{sec:currModel}

\begin{figure*}
\includegraphics[width=\largefig]{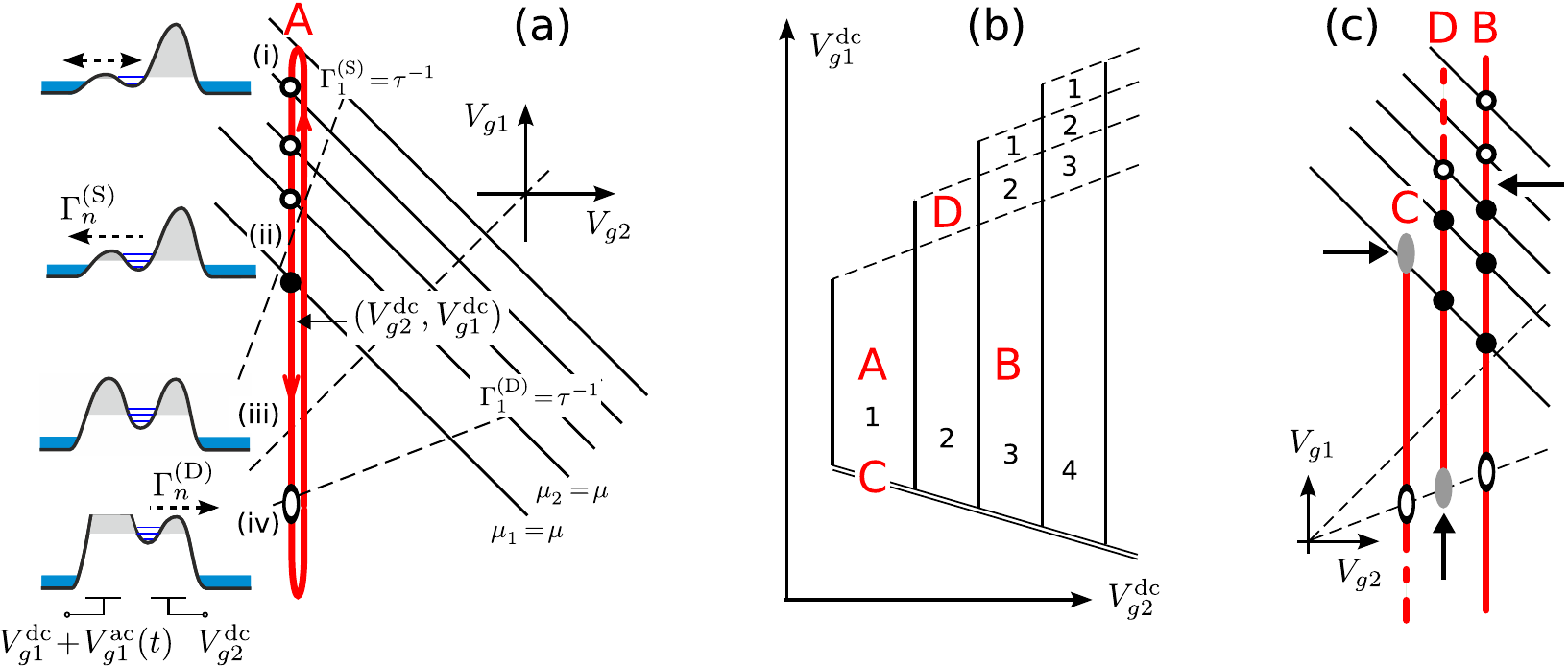}
\caption{Single-gate non-adiabatic pumping scheme for a tunable-barrier QD leading to robust quantization. (a)
The thick (red) path traces the instantaneous values of the gate voltages controlling the barriers to the source ($V_{g1}$) and the drain ($V_{g2}$) respectively; diagrams (i) to (iv) show schematically the real-space potential at specific time instants during the pumping cycle [cf.\ Figure~\ref{fig:barrier_lat}(c)].
(b) A typical two-dimensional map of the average pumped current, in units of electron charge per cycle, as function of the dc settings of the gate voltages with modulation amplitude and frequency fixed at sufficiently large values. The point $A$ corresponds to the particular example detailed in panel (a).  The quantized plateaux boundaries at the top of the panel (dashed lines) are set by incomplete  emission at phase (iv), the double line at the bottom marks the breakdown of the loading phase (i), the position of the vertical continuous lines is set by the outcome of the charge capture process (ii). (c) Particular pumping paths corresponding to points $B$, $C$ and $D$ in the pumping current map (b). Arrows mark the stage of the cycle which limits the number of transferred electrons.}
\label{fig:basicPrinciple}
\end{figure*}

A simple and robust pumping scheme which gives access to charge capture statistics
is the single-gate modulation of a tunable-barrier quantum dot shown in Fig.~\ref{fig:basicPrinciple};
the corresponding experimental realizations have been described in Section~\ref{sec:plungerBarrier} 
(see Fig.~\ref{fig:blumenthal}(b) and related discussion).
The voltage $V_{g1}$ on the entrance  gate is modulated periodically with a large ac amplitude, resulting in a sequence of confining potential configurations marked from (i) to (iv) in Fig.~\ref{fig:basicPrinciple}(a).
The connection to the charge capture statistics is established by identifying the generated dc current, 
$I=q_e f \langle n \rangle$, with the first moment,
 $\langle n \rangle = \sum_n n P_n$, of the probability distribution $P_n$ given by \eqref{eq:gradtofit} or \eqref{eq:deacyfull}.
Such simplification is justified only if the device is tuned into the appropriate operation regime.
The tuning has to rely on the pumping current $I(V_{g2}^{\rm dc},V_{g1}^{\rm dc})$ as function of the dc offset voltages on the gates; for large-amplitude harmonic modulation it shows a characteristic plateaux structure at integer levels of $\langle n \rangle$
shown schematically in Fig.~\ref{fig:basicPrinciple}(b). Quantization plateaus corresponding to Fig.~\ref{fig:basicPrinciple}(b) have been 
measured experimentally \cite{kaestner2008,kaestner2009a,kaestner2010c,wright2011,kataoka2011,Fletcher2012,seo2014}, see an example in Fig.~\ref{fig:excitedStatesKataoka}(a). 

The pumping contour for capture-dominated operation delivering one electron per cycle is shown in Fig.~\ref{fig:basicPrinciple}(a).
Multiple electrons are loaded on the dot in phase (i) when the entrance barrier is low enough for sufficiently long to establish near-equilibrium charge distribution on the dot. During the gradual decoupling phase (ii), the consequent top-electron levels $\mu_n$  emerge above the electrochemical potential of the source $\mu$, and some of the electrons escape back. The corresponding crossing points between the contour (thick red line) and 
the resonance lines  $\mu_n(V_{g2}^{\rm dc},V_{g1}^{\rm dc})=\mu$ (set of parallel solid anti-diagonal lines) are marked by open circles 
if the barrier is still open enough to allow electron escape, and by filled disks if the corresponding escape rate $\Gamma_n^{(S)}$ is too low for the backtunneling
to occur.  Eventually both the source and the drain barriers become sufficiently opaque to prevent any further change in the number of confined particles
(the isolation phase (iii) in Figure~\ref{fig:basicPrinciple}(a)). In the sudden decoupling limit,  the phase (ii) shrinks to a point 
separating the adiabatic loading (i) from the isolation (iii) phase of the pumping cycle.  After phase (iii), $V_{g1}$ keeps growing even more negative, and eventually enables electron escape into the drain once the corresponding rates, $\Gamma_n^{(D)}$, exceed the characteristic opening rate $\tau^{-1}$, as shown schematically by an ellipse marking the 
emission phase (iv) in Figure~\ref{fig:basicPrinciple}(a). In the second half of the pumping cycle the QD is returned back to the loading stage (i) through the same sequence of potential shapes  (iv) to (i). Capture of electrons from the drain is prevented by making sure that closing 
of the exit barrier happens when the first electron level on the QD is well above the Fermi sea in the drain.  

% In contrast to adiabatic pumps, the corresponding settings of the barrier-tuning gates can not be easily deduced from the
The operation scheme depicted in Figure~\ref{fig:basicPrinciple}(a) and marked by point $A$ in Figure~\ref{fig:basicPrinciple}(b) is robust against changes in the modulation amplitude and dc offset for $V_{g1}$
as long as the loading (i) and the emission (iv) stages take place properly. 
This can be seen from Figure~\ref{fig:basicPrinciple}(c) where several pumping trajectories with different
dc offsets are depicted.
The quantization plateaux boundaries along 
$V_{g1}^{\rm dc}$ axis are set either by insufficient loading (case $C$ in Figures~\ref{fig:basicPrinciple}(b) and (c)) or
incomplete emission (case $D$). 
For a larger ac amplitude $V_{g1}^{\rm ac}$, a larger shift in $V_{g1}^{\rm dc}$ would be  needed to  
turn a loading-limited trajectory into an emission-limited one, hence the length of the quantization plateaus along  $V_{g1}^{\rm dc}$ 
(solid vertical lines in Figure~\ref{fig:basicPrinciple}(b))
grows with increasing modulation amplitude \cite{kaestner2008}. Additional steps 
at the top  of Figure~\ref{fig:basicPrinciple}(b), such as the one corresponding to case $D$, are due to  emission-rate separation between different charge states 
at stage (iv). Identification of plateaux edges and their connection to specific phases of a non-adiabatic pumping cycle can be done along similar lines for other choices of control voltages, see \cite{fujiwara2008,miyamoto2008,YamahataPRB2014,tettamanzi2014,Yamahata2014}. Deliberately tuning the pump into emission- or loading-limited regimes has been used to explore voltage- and temperature- dependence of the relevant charge exchange rates \cite{miyamoto2008,FletcherPRL2013,Yamahata2014}.

Pumping trajectories $A$ and $B$ are both capture-limited and allow to optimize the average number of captured electrons
$\langle n \rangle$ by tuning  $V_{g2}^{\rm dc}$ and thus shifting the position of the decoupling phase (ii) relative to the resonance lines 
$\mu_n =\mu$. $\langle n \rangle$ is set by the number of out-of-equilibrium charge states for which relaxation back to the source is blocked, ie.\ the number of the crossing points marked by filled disks in Figures~\ref{fig:basicPrinciple}(a) and (c).  
In terms of the theory described in Section~\ref{sec:modIntro}, the filled-disk crossings are associated with  
negligible integrated escape rates, $X_n \ll 1$, and essentially decoupled charge states by the time of the crossing, $\tilde{\mu}_n < \mu$, whereas the open circles correspond to $n$ with $X_n \gg 1$ and $\tilde{\mu}_n > \mu$.
Tuning $V_{g2}^{\rm dc}$ more positive (e.g., going from $A$ to $B$)  reduces the backtunnelling rates 
and shifts the onset of backtunnelling to later times (when $V_{g1}(t)$ is more negative), in both ways reducing $X_n$ and making the energies levels at decoupling $\tilde{\mu}_n$ more negative. 
These effects correspond to moving the crossing points corresponding to phase (ii) down the resonance lines $\mu_n=\mu$ (black anti-diagonals) and up the level-lines of $\Gamma_{n}^{(S)}$  (e.g., the upmost dashed line in  Fig.~\ref{fig:basicPrinciple}(a)). The transitions of $\langle n \rangle$ from $n_0$ to $n_0+1$ (an open circle turning into a filled one) happen when $X_{n_0} \sim 1$ and $\tilde{\mu}_{n_0} \sim \mu$, in accord with the limiting forms of the capture probability
distribution \eqref{eq:deacyfull} and \eqref{eq:gradtofit}.

Connecting the shape of the current quantisation steps $I(V_{g2}^{\rm dc})$ under capture-dominated conditions with the universal distributions discussed in Section~\ref{sec:modIntro} 
requires the knowledge of parametric dependence for the rates and energies during the capture phase (ii). Linear effect of gating on electron energies and exponential effect on charge exchange rates motivates the following functional dependencies:
%\numparts
\begin{eqnarray} \label{eq:fittingansatz} 
   \ln X_n & = & -\alpha_n^{X} V_{g} + \Delta_n^{X} \, , \\
   \tilde{\mu}_n & = & -\alpha_n^{\mu} \, V_{g} + \Delta_n^{\mu} \, , \label{eq:fittingmu}
\end{eqnarray}
%\endnumparts
where $\alpha_n^{X}$, $\Delta_n^{X}$, $\alpha_n^{\mu}$, and $\Delta_n^{\mu}$ are constants, 
and $V_{g}$ is a dc voltage affecting  the conditions of charge capture ($V_g \to V_{g2}^{\rm dc}$ for the present example). The unknown linearization parameters in the r.h.s.\ of \eqref{eq:fittingansatz} and 
\eqref{eq:fittingmu}
 can be either treated as phenomenological constants \cite{fujiwara2008,kaestner2009a,kaestner2010a,Fricke2013}
or calculated from the electrostatics \cite{YamahataPRB2014} or microscopic modelling \cite{Fletcher2012} under the assumptions outlined in Section~\ref{sec:modIntro}.
%The decoupling phase (ii) has been model in Section~\ref{sec:modIntro} where the initial, $t_0$, and final, $t^{\ast}$, time moments correspond to the loading  (i) and isolation (iii) stages, respectively. The number $\langle{n}$
For example, parametric pumping with tunneling rates $\Gamma_{n}^{(S)} \propto \exp [-(E_b-\mu_n)/\Delta_b]$ and equal and linear plunger effect of both gates, $\partial \mu_n/\partial V_{g1}\!=\!\partial \mu_n/\partial V_{g2}\!=\!{\rm const}$, would result in 
$\alpha^{X}_n = (|\partial E_b/\partial V_{g1}| -|\partial E_b/\partial V_{g2}|)/\Delta_b$ and $\alpha^{\mu}_n= \Delta_{\rm ptb} \, \alpha^{X}_n$ independent of $n$. However, energy-dependence of the entrance barrier sharpness parameter $\Delta_b$ \cite{Giblin2013} or additional conductance paths (e.g, charge traps or isolated donors) may lead to 
$n$-dependent $\alpha_n$'s.

\begin{figure}
\includegraphics{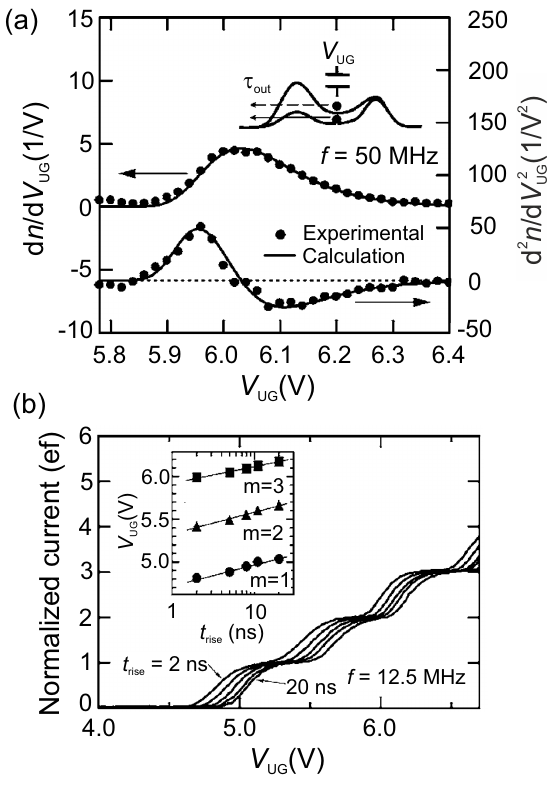}
\caption{(a) Calculated first and second derivatives of the pump-current fitted to experimental data (single quantization step)~\cite{fujiwara2008}. (b) Dependence of pump-current characteristic on rise time. The inset shows the corresponding shift of the current step positions~\cite{fujiwara2008}. Reproduced with permission of the authors. Copyright 2008, AIP Publishing LLC.}
\label{fig:Fujiwara}
\end{figure}

Gate-voltage- and rise-time-dependence of $X_n$ for non-adiabatic capture-limited charge pumping has been investigated in the pioneering work of
Fujiwara~\emph{et al.}~\cite{fujiwara2008}. 
The device was realized in a silicon nanowire MOSFET, similar to that shown in Figure~\ref{fig:BarrMod_Devs}(b). 
The dot is defined between the gates G1 and G2, to which a pulse-modulated voltage $V_{G1}$ and a fixed voltage $V_{G2}$ have been applied, respectively. 
The rise time $t_\mathrm{rise}$ is defined as the time to switch between $V_{G1H}$ (high) and $V_{G1L}$ (low). During the high-state the dot equilibrates with source (stage (i) in terms of our schematics). 
Electrons are emitted to the drain during the low-state of $V_{G1}$ (emission stage (iv)).
The voltages setting $V_{G1L}$ and $V_{G1H}$, as well as the dc voltage $V_{\rm UG}$ on the upper gate have been
tuned to make sure that loading and emission take place with sufficient fidelity, and the total current $I=q_e f \langle n\rangle$ is a measure 
of the capture statistics during during the pulse rise from $V_{G1H}$ to $V_{G1L}$. 
% (corresponding to combined stretch (ii)+(iii) of the pumping trajectory in Fig.~\ref{fig:basicPrinciple}(a)). 
The measurement temperature was 20~K.

The rise time $t_\mathrm{rise}$ controls the duration of the capture phase with the decoupling time $\tau$ proportional to $t_\mathrm{rise}$ under the assumption of adiabatic rates (see Section~\ref{sec:modIntro}). Voltage $V_\mathrm{UG}$, applied to the upper gate (see Figures~\ref{fig:BarrMod_Devs}(b) and \ref{fig:Fujiwara}(a)), is used to tune the depth of the confining potential well and hence the electron escape rate $\tau_{\rm out}^{-1}$ during the capture phase ($\Gamma^{(S)}_n$ in our notation). The measured dependence of  $\langle n \rangle$ on $V_\mathrm{UG}$ is shown in Figure~\ref{fig:Fujiwara}. The inset in Figure~\ref{fig:Fujiwara}(b) marks position of the peaks in $d\langle n \rangle/dV_\mathrm{UG}$ for $t_\mathrm{rise}=2 \ldots 20 \, \rm{ns}$.
The model calculation
reported in Figure~\ref{fig:Fujiwara}(a) is  as single-step fit to $\langle n \rangle =e^{-X(V_{\rm UG})}$ with 
$X = \exp ({- \alpha V_{\rm UG} + \ln t_{\rm rise}+\rm{const}})$. The fitted value of $\alpha =(79 \, \rm{m}e\rm{V})^{-1}$ corresponds
very well to the value of the slope $\Delta \ln t_{\rm rise}/\Delta V_{\rm UG}= (82 \, \rm{m}e\rm{V})^{-1}$ for the straight lines in the inset of 
Figure~\ref{fig:Fujiwara}(b). These lines correspond to $X_m = \rm{const}$ for $m=1,2,3$  (see \eqref{eq:fittingansatz} with $V_{g}\to V_{\rm UG}$ and equal $\alpha_m^{X}=\alpha$) confirming exponential voltage dependence and the parametric nature of the escape rate modulation under the conditions of the experiment.

%\commVK{Decay cascade theory as the argument why $P_0$ and $P_2$ can be trusted at the same time.}
For the purpose of achieving most accurate quantization, the primary goal for modeling is the current quantization plateaux, on which 
at least three components of the probability distribution $P_n$ contribute. 
The decay cascade model described in Section~\ref{sec:modIntro} provides   
a robust fitting formula for the average current in the limit of large plunger-to-barrier ratio $\Delta_{\rm ptb}$. Eqs.~\eqref{eq:deacyfull} and \eqref{eq:fittingansatz} with $\alpha^{X}_n=\alpha$ give \cite{kaestner2010a}
\begin{eqnarray} 
 I(V_{g}) & = & q_e \sum_m m P_m  \nonumber \\
  & = & q_e f \sum_{m=1}^{N_{\rm max}} \exp\left[ -\exp\left( -\alpha V_{g} + \Delta_m^X \right)\right], \label{eq:decayFit2}
\end{eqnarray}
where $\alpha$ and $\Delta_m^X$ are the fitting parameters. 

On the first plateaux in a sequence of well-defined steps in $I(V_{g})$, $P_1$ is close to one,
and the  probabilities of keeping an extra electron ($P_2$) or missing one ($P_0$) can be combined into the total error probability per cycle $P_{\rm err} = 1- P_1 \approx P_0+P_2 \ll 1$. Fitting the first plateaux  to \eqref{eq:decayFit2} with $N_{\rm max}=2$ and extracting
the parameter $\delta_2 =\Delta_2^X-\Delta_1^X$ gives an figure-of-merit directly related to the minimal $P_{\rm err}$ \cite{kaestner2010a}.
Analytically, the minimum of $P_{\rm err}$ can be estimated from $P_0(V_{g})=P_2(V_{g})$ as 
\begin{equation}\label{eq:Perrest}
\min P_{\rm err} \approx 2 \, \delta_2 \exp(- \delta_2) \, ,
\end{equation} with the accuracy of the estimate better than 20\%  for $\delta_2 > 10$.

\begin{figure*}
\includegraphics[width=\largefig]{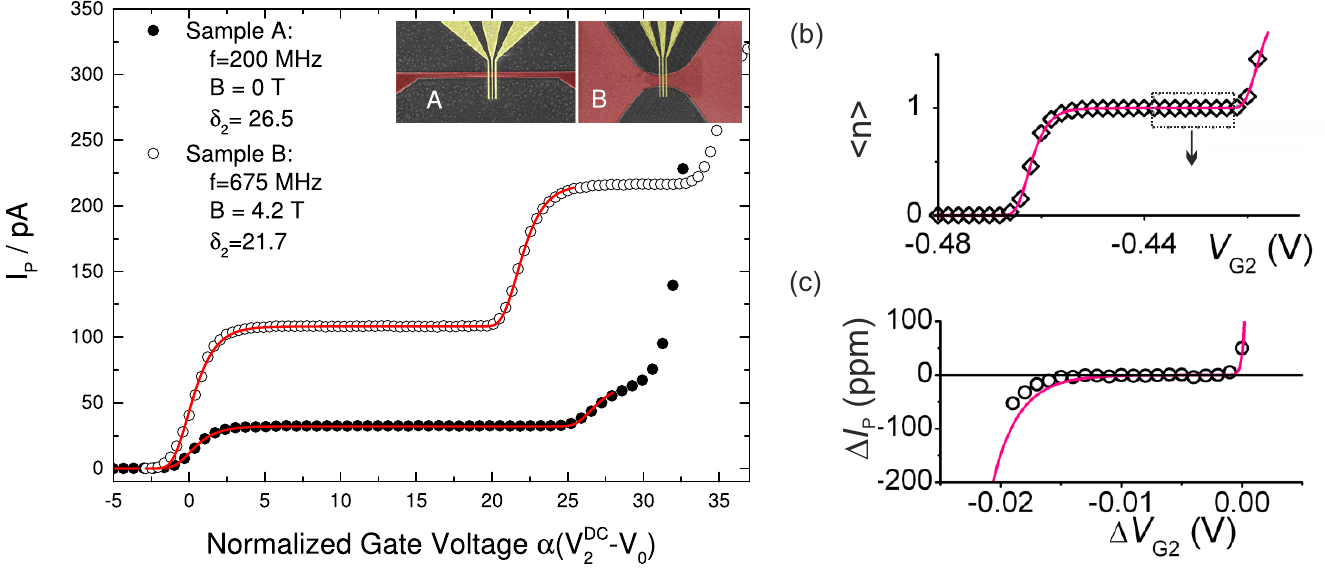}
\caption{(a) Example of current traces of single-gate operated pumps, fitted to the decay cascade model. The pumps differ in their channel geometry, where the tapered channel optimizes operation in magnetic field environments~\cite{Leicht2012}. Data are plotted on an offset gate voltage scale, normalized by the fitting parameter $\alpha$ from \eqref{eq:decayFit2} (b) Current plateaux obtained from a low resolution measurement, fitted to the decay cascade model~\cite{giblin2012}. Same current plateaux but using high-resolution data compared to the fit in (b) (line) for the range indicated and plotted on an offset gate voltage scale. Reprinted by permission from Macmillan Publishers Ltd: Nature Communication \cite{giblin2012}, copyright 2012.
\label{fig:decayVeri}}
\end{figure*}

The potential to estimate minimal achievable quantization error from the shape of the current steps has
made the decay cascade model \eqref{eq:decayFit2} a popular tool for analysis of non-adiabatic quantized charge pumps 
\cite{kaestner2010a,giblin2012,Fletcher2012,kaestner2012,Rossi2014,seo2014}. 
Using $\delta_2$ as an easily accessible figure of merit has enabled phenomenological exploration of different optimization strategies for non-adiabatic pumps; examples are covered in Sections.~\ref{sec:magneticExp} and \ref{sec:waveForm} below. 
This approach allows comparison of pumps beyond the limits of measurement uncertainty, an example  is shown in Fig.~\ref{fig:decayVeri}(a)~\cite{Leicht2012}. The extracted $\delta_2$ values of %\sout{26.5} 
27 and 
%\sout{21.7} 
22 can be related to the relative deviation of the modeled current \eqref{eq:decayFit2} from the ideal value of $1 \, q_e f$ at the flattest part of the plateau~\cite{kaestner2010a} and would in this case be $10^{-10}$ and $10^{-8}$, respectively (very close to the analytic estimate of $P_{\rm err}$ given by \eqref{eq:Perrest}).
However, such extrapolations need to be treated with great caution given the number of difficult-to-verify assumptions leading to the decay cascade fitting formula \eqref{eq:decayFit2}.

The shape of the current quantization plateaux has been investigated by high resolution measurements by Giblin~\emph{et al.}~\cite{giblin2012} and compared with the decay cascade model. Figure~\ref{fig:decayVeri}(b) shows the average number of pumped electrons $\qav{n} = I /(q_e f)$ as a function of drain barrier voltage $V_{G2}$ for a low resolution measurement. The pump frequency has been set to 945$\,$MHz using an optimized pulse shape (see Section~\ref{sec:waveForm}). The fit to \eqref{eq:decayFit2} is shown by the red line. The corresponding high-resolution measurement can be seen in (c). The double-exponential shape has clearly been reproduced, and deviations become visible at high resolution. 

In retrospect, the ubiquity of the current quantization steps that fit well to the decay cascade model in \emph{single-gate} pumps
comes at no surprise: ensuring complete emission by driving the same gate which controls capture requires a large amplitude modulation and a strong plunger function for the gate, hence large $\Delta_{\rm ptb}$. At $\Delta_{\rm ptb} > kT$ the steps are asymmetric \cite{Kashcheyevs2012a} and at
$\Delta_{\rm ptb} \gg kT$ the decay cascade limit is justified (see Section~\ref{sec:modIntro}). In contrast, \emph{two-gate} operation in a turnstile 
mode with high-fidelity tunable barriers \cite{jehl2013,YamahataPRB2014,Rossi2014} allows keeping the QD energy levels largely constant with respect 
to the leads, and thus makes the temperature-dominated limit of sudden decoupling more easily accessible (see discussion of \cite{YamahataPRB2014} 
in Section~\ref{sec:CountStatModel}).

\subsection{Shot noise}
\label{sec:noise}
The measurement of the current noise power spectrum $S_I(f_0)$ as function of frequency $f_0$ provides an experimental proof 
of quantized charge pumping independent of the average current value. 
In the  low-frequency limit of $f_0 \ll f$, where $f$ is the pumping frequency, the current noise power $S_I(f_0) \to S_I^0$ is expected to become frequency-independent, and reflect the dispersion in the number of electron transferred per cycle~\cite{Galperin2001,Robinson2002},
$S_I^0 = 2 \, e^2 f (\langle n^2 \rangle-\langle n \rangle^2)$.  For an ideal quantized charge pump it becomes zero, while a non-zero noise power $S_I^0$ reveals directly the pumping errors: $\langle n^2 \rangle-\langle n \rangle^2 \approx P_{\rm err}$ when the probability $P_{\rm err}$ of delivering a wrong number of electrons per cycle is small.
By combining measurements of the shot noise and the average current, a missing-electron error can be distinguished from delivering an extra one, e.g. separating the two contributions to $P_{\rm err}=P_0+P_2$ near $\langle n \rangle \approx 1$ even if $P_0$ partially compensates $P_2$ in the average current \cite{Robinson2002}.

Shot noise measurements have been carried out for single-gate-driven pumps by Maire~\emph{et al.}~
\cite{maire2008,Maire2009}
at operation  frequency of $f=400 \,$MHz with the noise power averaged over the range of $f_0 = 5$\ldots$15 \,$ kHz.
The results have confirmed association of plateaus in the average current with low error probabilities, and upper limit on the minimal $P_{\rm err}$ was estimated  to be $4 \%$.

\begin{figure}
\includegraphics[%trim=0 60 0 50, clip, 
width=\smallfig]{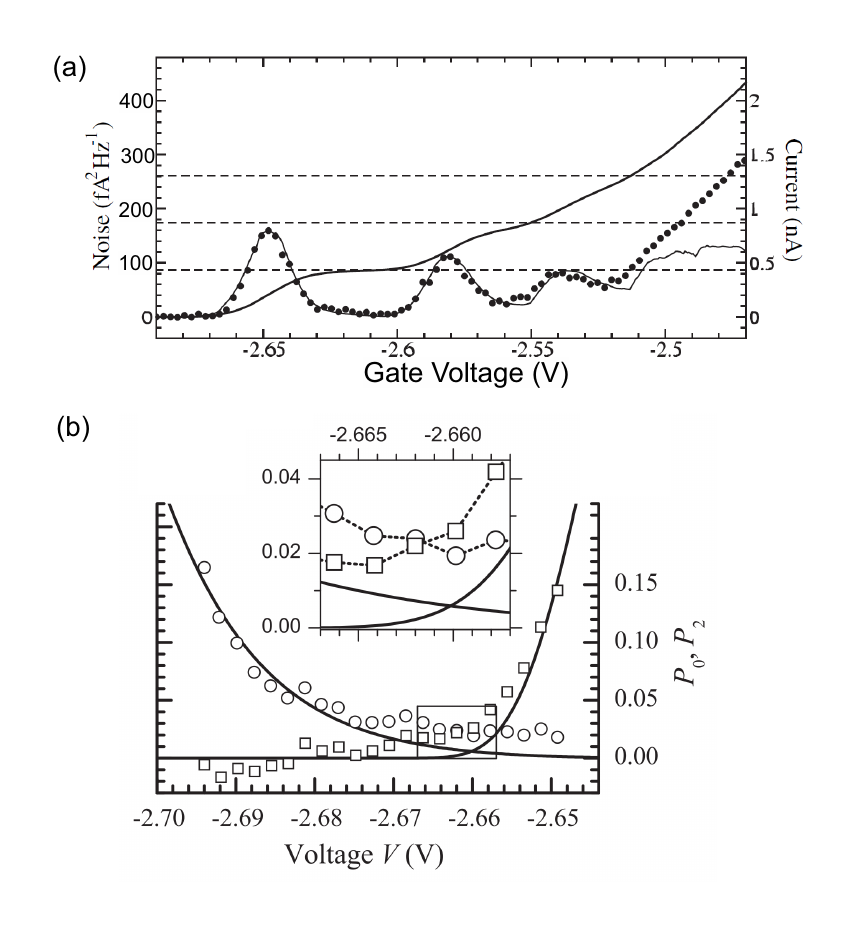}
\caption{
(a) SAW-driven current (bold line) and current noise (dots) varying with gate voltage. Thin line shows a model calculation combining switching-type and shot noise. Reprinted with permission from \cite{Robinson2005}. Copyright 2005 by the American Physical Society. (b) Comparison of decay cascade model~\cite{kaestner2010a} (solid line) with probabilities $P_0$ ($\bigcirc$) and $P_2$ ($\square$) extracted from the shot-noise measurements in \cite{Robinson2005}.
\label{fig:RobinsonNoise}
}
\end{figure}

Robinson~\emph{et al.} have performed a series of experiments measuring the shot noise in SAW-driven pumps~\cite{Robinson2002, Robinson2005} (see Section~\ref{sec:SAW}). The pump was operated at 2.7$\,$GHz and the shot noise was measured at a relatively high frequency above $\approx 1\,$MHz \cite{Robinson2005}. In a previous experiment at $\approx$ 1$\,$kHz the measured noise level was nearly 3 orders of magnitude above the theoretical value~\cite{Robinson2002}, which was interpreted as being caused by switching the charge states of single-electron traps close to the 1D channel. However, the noise determined from the high-frequency range is suppressed on the current plateaus and reaches values on the order of $100\,$fA$^2$Hz$^{-1}$ at the transitions between the plateaus, as seen in Fig.~\ref{fig:RobinsonNoise}(a). The analysis shows that close to the quantized value the noise is dominated by shot noise whereas away from this range the noise mostly arises from switching the charge states of electron traps.
The probabilities $P_0$ and $P_2$ extracted from the data on the first quantization plateaux are shown in Fig.~\ref{fig:RobinsonNoise}(b).

Several theoretical models of SAW-driven charge pumping \cite{Aizin1998,Flensberg1999,maksym1,Robinson2001} suggest it is capture-limited at large driving amplitudes and frequencies.  Characteristic asymmetry of current quantisation steps suggests that decay cascade model can be used to describe the parametric dependence of $P_{n}$ \cite{kaestner2010a}.
Fitting \eqref{eq:decayFit2} to the average current data from \cite{Robinson2005} gives separate estimates of $P_0$ and $P_2$ that can be compared to the experimentally extracted ones, see solid lines in  Fig.~\ref{fig:RobinsonNoise}(b). Observed qualitative agreement suggests that the decay cascade model is a reasonable quantization benchmark for SAW-driven charge pumps.

\subsection{Error rates by electron counting}
\label{sec:CountStatModel}

\begin{figure}
\includegraphics{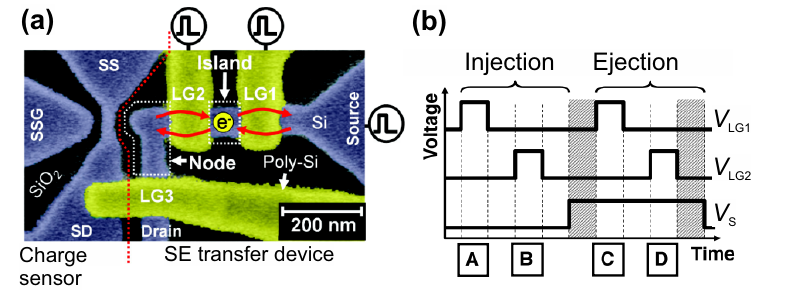}
\caption{(a) False-colour image of the device employed in \cite{YamahataPRB2014} before upper gate deposition. (b) Pulse sequence to obtain shuttle transfer to and from the node. Hatched regions are 4 orders of magnitude longer than the other time intervals. Reprinted with permission from \cite{YamahataPRB2014}. Copyright 2014 by the American Physical Society.}
\label{fig:Yamahata}
\end{figure}

Keeping track of individual electron transfer events by on-chip components in addition to the pump itself enables additional insight into the pumping mechanisms and a model-independent way to 
ascertain accuracy of quantization.

For Si-based tunable-barrier turnstile devices (see Sections~\ref{sec:BarrierMod} and \cite{fujiwara1}), Yamahata~\emph{et al.}~\cite{yamahata2011,YamahataPRB2014} have performed error counting by shuttling single electrons between a lead and a charge-accumulating node and detecting in real time the number of electrons on the node. Fig.~\ref{fig:Yamahata}(a) shows the turnstile device and the charge sensor to the right and the left sides of the dotted line, respectively. Shuttling is achieved by applying the pulse sequence shown in Fig.~\ref{fig:Yamahata}(b) to the gates LG1 and LG2, and to the source contact S. The error rate of the single-electron capture process was determined to be as low as 100 ppm. Considering the rise times of the voltage pulses (2\, ns) the authors suggest that quantized currents at 100MHz can be generated at this error rate.

The authors of \cite{YamahataPRB2014} have used a model of sudden decoupling from thermal equilibrium (see Section~\ref{sec:modIntro}) to establish a connection between the results of  charge counting and the measurements of the average current done separately on the same device in a continuous  turnstile-like operation mode.
 Fitting a sequence of symmetric current steps to a model equivalent to  \eqref{eq:gradtofit} and \eqref{eq:fittingmu} they have obtained an estimate of the addition energy $E_{\mathrm{add}}$.
%where %$E_{\mathrm{add}} = \tilde{\mu}_n- \tilde{\mu}_{n\!-\!1}$ 
The corresponding minimal $P_{\rm err} =2 P_{n_0\!-\!1} =2 P_{n_0\!+\!1}$ can be estimated from \eqref{eq:gradtofit}  at optimal $\mu = \tilde{\mu}_{n_0}-E_{\mathrm{add}}/2=\tilde{\mu}_{n_0\!-\!1}+E_{\mathrm{add}}/2$,  
\begin{eqnarray}
  \min P_{\rm err} & \approx & 2 \, e^{-E_{\mathrm{add}}/(2 kT)} \, . 
\end{eqnarray}
The value of $E_{\mathrm{add}}/kT$ extracted from the current measurements gives  $\mathrm{min} \, P_{\rm err} \approx 24 \, \mathrm{ppm}$ 
% \label{eq:Yamahata}
which is smaller than $100 \, \mathrm{ppm}$ obtained in the shuttle error measurements.
This discrepancy has been discussed in \cite{YamahataPRB2014} in terms of charging effects in the node  which may shift the operation point away from the optimum \cite{fricke2011}.

\begin{figure}
\includegraphics{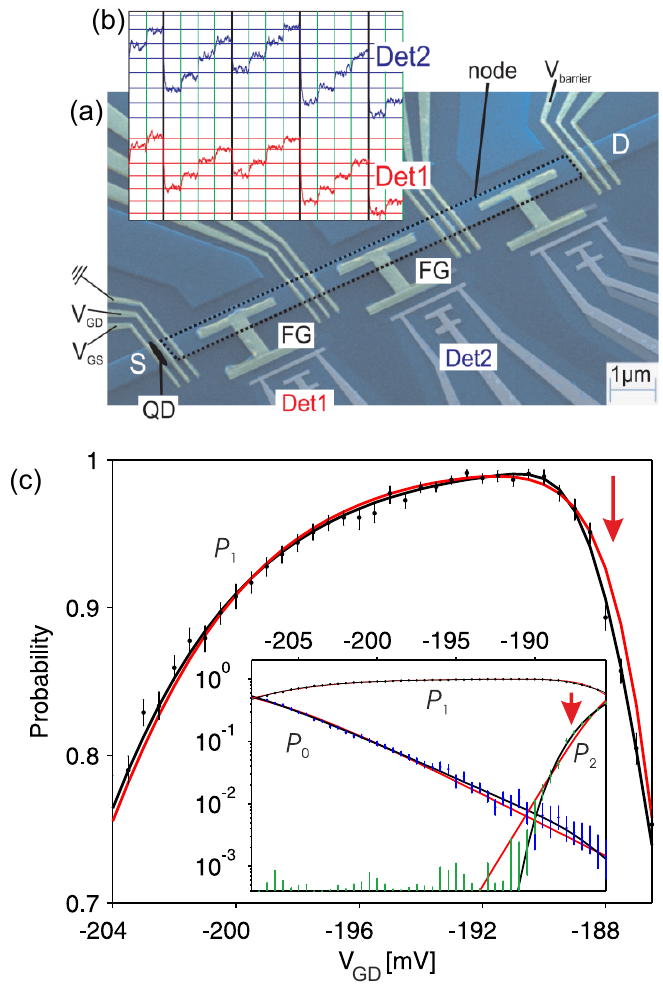}
\caption{(a) Image of the device used for determining the counting statistics~\cite{Fricke2013}. The upper half shows the semiconducting part consisting of an 800 nm wide channel (light blue) crossed by topgates (yellow). The QD is formed by the leftmost group of the topgates, between source (S) and drain (D). The light-grey parts in the lower half form the SETs used for single-charge detection. (b) Detector signals of the  charge transfer sequence. (c) Measured probabilities $P_1$ as function of $V_\mathrm{GD}$, compared to different theories. Inset shows the full distribution on a logarithmic scale.
\label{fig:decayVeri2}}
\end{figure}

Electron capture in a QD has been explored by Fricke~\emph{et al.}~\cite{Fricke2013} using 
a GaAs-based device shown in Figure~\ref{fig:decayVeri2}(a). A set of gates is used to create barriers that define the QD (between $V_\mathrm{GS}$  and $V_\mathrm{GD}$, marked ``QD'' in the figure) and a larger node (between $V_\mathrm{GD}$  and $V_\mathrm{barrier}$, marked by a long dashed rectangle).
Below the node two single-electron transistors (SET) based on Al--AlO$_x$--Al tunnel junctions are placed as detectors (Det1 and Det2) for the charge on the node. The SETs are operated at fixed voltage bias, using the current as detector signal.

%Capturing electrons in the QD from source (S) is achieved by modulating the gate voltage  and keeping $V_\mathrm{GD}$ fixed at the pinch-off value. During the first part of the modulation cycle the QD decouples from the electron reservoir, i.e. capturing a certain number of electrons in the isolated QD. In the second part of the cycle emission over the opposite barrier (defined by $V_\mathrm{GD}$) is induced, since the GS-barrier rises further while at the same time the transmission through the GD-barrier increases due to the punger-barrier coupling (see Section~\ref{sec:plungerBarrier}). Emission of all captured electrons to the node between QD and the barrier defined by the gate voltage $V_\mathrm{barrier}$ has been ensured.

The QD is driven through a sequence of steps from loading (i) to emission (iv) by modulating $V_\mathrm{GS}$ as depicted schematically in Figure~\ref{fig:basicPrinciple}.  Consecutive pump cycles charge up the node and the resulting change in the node potential after each cycle is detected, as shown in Figure~\ref{fig:decayVeri2}(b). The node charge is reset every three cycles by switching $V_\mathrm{barrier}$ and equalizing the potentials of the node and the drain $D$. In this way the probability $P_n$ to capture $n$ electrons form the source S could be resolved as function of $V_\mathrm{GD}$ for $n$ up to $4$. Figure~\ref{fig:decayVeri2}(c) shows $P_1$ in the main plot for which an initialization probability of 99.1$~\%$ has been achieved at $V_\mathrm{GD} \approx -192.5\,$mV. The inset shows $P_0$, $P_1$, and $P_2$ on a logarithmic scale for $V_\mathrm{GD}$ in the region where single-electron capture dominates.

%\begin{eqnarray} % \label{eq:gradtofit}
%   P_n & = \left[1- f\bm{\left(}\tilde{\mu}_{n\!+\!1}^{}\bm{\right)}\right] \prod_{m=1}^{n} f\bm{\left(}\tilde{\mu}_m\bm{\right)} \, ,
%\end{eqnarray}
%with $f(x) = 1/(1 + e^x)$, and $\tilde{\mu}_n = \alpha_{\mu, n} V_\mathrm{GD} + \Delta_{\mu, n}$, with $\alpha_{\mu, n}$ and $\Delta_{\mu, n}$ fitting %parameter. The probability distribution of the athermal limit is given by
%\begin{eqnarray}
%P_n & = e^{ -X_{n}} \prod_{j=n+1}^{N} \left( 1 - e^{-X_{j}} \right) \, ,
%\end{eqnarray}
%with $\ln X_n = -\alpha_{X, n} V_\mathrm{GD} +\Delta_{X, n} \,$, and $\alpha_{X, n}$ and $\Delta_{X, n}$ fitting parameters. 

Parametric dependence of $P_{n}(V_\mathrm{GD})$ has been compared to the two extremes of charge capture
described in Section~\ref{sec:modIntro} using \eqref{eq:gradtofit}, \eqref{eq:fittingmu} for the generalized grand canonical (sudden decoupling limit, red line in Figure \ref{fig:decayVeri2}(c)) and \eqref{eq:deacyfull}, \eqref{eq:fittingansatz} for the decay cascade (gradual decoupling limit, thick black line) distributions, respectively. The decay cascade distribution makes a better fit to the observed shape of probability distribution, even if additional flexibility is allowed by $n$-dependent parameters $\alpha_n^{\mu}$ and  $\alpha_n^{X}$, see regions marked by arrows in Figure~\ref{fig:decayVeri2}(c). These results suggest that the large plunger-to-barrier ratio $\Delta_{\rm ptb}$, not the frozen equilibrium fluctuations, are limiting the precision of quantization in the GaAs-based realisation of \cite{Fricke2013} studied at $T = 25 \,$ mK.

Further development of electron counting techniques in the context of metrological applications is discussed in Section~\ref{sec:erroraccounting}.

\section{Optimization towards higher precision}
\label{chap:metrology}
The discussion so far has mainly dealt with the physics and the technology of tunable-barrier quantized pumps, leaving out their potential applications in the field of metrology. An important feature of any quantized charge pump is their current output being traceable directly to the elementary charge $e$. Provided a sufficient precision they can play an important role in the ongoing process of restructuring the International System of Units (SI)~\cite{mills2006, mise2009, milton2010a}. Therefore much effort has been devoted to improving their precision, which this section focuses on. In Section~\ref{sec:metro} we will first provide more details on the metrological relevance, before moving on to specific approaches to achieve higher precision in Sections~\ref{sec:magneticExp},  \ref{sec:internalExcitations} and \ref{sec:waveForm}. Finally,  Section~\ref{sec:erroraccounting} reviews first experimental results on a method which reduces the uncertainty in the current output beyond the stochastic error of the pump by detecting and processing individual error events.

%%%%%%%%%%%%%%%%%%%%%%%%%%%%%%%%%%%%%%%
\subsection{Precision requirements for metrological relevance}
\label{sec:metro}

Absolute measurements can fundamentally not be more precise than the uncertainty of the realization of the corresponding unit. Therefore, the realization of a unit according to its definition with  smallest possible uncertainty is a permanent challenge in the field of metrology. Although the existing SI, the international system of units, fulfills widely the requirements of science and technology it is still far from the ideal of being \emph{available worldwide and stable for all times}. Therefore the metrological community has adopted the long term goal
%~\cite{mise2009} 
of basing all SI units on the invariants of nature --- the fundamental physical constants or properties of atoms \cite{CGPM2011}. In order to redefine units accordingly, several physical constants will be assigned exact values, including the elementary charge $e$. The challenge is now to design a procedure to obtain units from constants with as little error as possible. The units \emph{meter} and \emph{second} have been the first highly successful outcomes of this procedure \cite{SI2006a}.

% Realization of a NEW ampere
For the above reasons the present definition of the unit \emph{ampere}~\cite{SI2006} is considered problematic, in particular because it is linked, via a current-induced force, to the artifact-based kilogram. On the other hand, single-electron current sources transporting a charge $Q_S$ with a frequency $f$ can generate currents of $I = Q_S f$ with uncertainties in the ppm-range~\cite{giblin2012} and even less for lower frequencies~\cite{keller1996}. Since the frequency $f$ could be measured with atomic clocks of very low uncertainty and high stability, the ampere would be traced to the elementary charge, to which one then has to assign an exact value. However, there is a trade off between reliability in manipulation and frequency. The uncertainty of a device realizing a new definition should improve on the uncertainty of the realization of the existing unit.
Using the present ampere realization and corresponding experiments  that trace the ampere to the SI unit of the force  an uncertainty of around $10^{-7}$ has been achieved~\cite{sienknecht1986}.
 Moreover, this level of uncertainty is required for currents in the microampere range or higher to make practical current metrology feasible, as manifest from the calibration and measurement capabilities (CMCs) published in the BIPM Key Comparison Data Base. On the other hand, current scaling techniques using cryogenic current comparators (CCCs) have been designed for nanoampere currents (or higher) to be scaled up to the microampere range without degrading this uncertainty~\cite{steck2008}. From this one can see that at least nanoampere levels ($f \approx 5\,$GHz) with an uncertainty below $10^{-7}$ are required for the current standard to be of practical relevance. However, much lower currents may be sufficient owing to the recent development of the \emph{ultrastable low-noise current amplifier} --- ULCA \cite{drung2014}. It is a non-cryogenic instrument based on specially designed operational amplifiers and resistor networks. A CCC is also required to calibrate the current gain, but at much higher current levels. The calibration remains stable so that 1000-fold scaling at a level of $10^{-7}$ for currents of the order of $100\,$pA may be possible for the duration of several days.

% But first we need the NEW ampere; Following shows how pumps help to get the new ampere by QMT
An already established method for realizing the new definition utilizes Ohms law and combines the Josephson and quantum Hall effects realizing the volt and ohm, respectively. A crucial aspect for the application of the Josephson and the quantum Hall effects is the assumption that the fundamental relations $K_J = 2 e/h$ and $R_K = h/e^2$ are exact, with the Josephson constant $K_J$ and the von Klitzing constant $R_K$. Enhancing experimental confidence in these assumptions is still an ongoing goal in the field of modern fundamental metrology, and its need has been repeatedly emphasised by the international Committee on Data for Science and Technology (CODATA)~\cite{mohr2008}. One approach that has been intensively investigated is to realize the closure of the so called \emph{quantum metrological triangle} (QMT). In one version of a QMT experiment, dc voltage $U_J$ is obtained from frequency $f_J$ through the Josephson effect, $U_J = n f_J/K_J$. Dc current can then be derived using the quantum Hall effect, $U_J \times m /R_K$. Direct realization of dc current from frequency $f_\mathrm{SET}$ using single-electron currents sources, $I_\mathrm{SET} = Q_S f_\mathrm{SET}$ would be the third leg of the triangle. Here also $Q_S$ has been considered a phenomenological constant, differentiating the charge quanta in solid-state devices and electron charge in vacuum~\cite{piquemal2000, keller2009}. Equalizing the currents derived via the different paths results in $K_J R_K Q_S = m n f_J/f_\mathrm{SET}$. The result on a QMT experiment can thus be expressed as $f_J/f_\mathrm{SET} = 1 + \Delta_{\mathrm{QMT}} \pm u_\mathrm{QMT}$, where $m=n=1$, $\Delta_{\mathrm{QMT}}$ is the measured deviation from the expected equality of the two frequencies and $u_\mathrm{QMT}$ is the relative standard uncertainty attributed to the result.

The impact of QMT results can be classified according to the following uncertainty ranges $u_\mathrm{QMT}$ ~\cite{keller2009, scherer2012a, scherer2013}, where all quoted uncertainties in this section will be 1$\sigma$. The closure with $u_\mathrm{QMT}$ of about 1 part in $10^6$ has to be interpreted in terms of deviations of $Q_S$ from $e$. Uncertainties in the range of 5 parts in $10^7$ to 2 parts in $10^8$ would have an impact on  $Q_S$ and $K_J$, while even lower uncertainties would bear on the correction for all three quantum electrical effects. A milestone QMT experiment has been carried out by Keller~\emph{et al.}~\cite{keller1999,keller2009} using pumps based on a series of small metallic islands with fixed tunnel barriers. No evidence for a deviation has been found up to a relative uncertainty of $9 \times 10^{-7}$ \cite{keller2007a}. A similar experiment is currently being pursued by Scherer~\emph{et al.} with the potential to achieve relative uncertainties of about $2 \times 10^{-7}$~\cite{scherer2012a, scherer2013}.

Due to the rapid progress in tunable barrier pumps they are now also considered potential candidates for QMT experiments. Optimizing gate drive waveforms of a tunable barrier pump and applying a magnetic field have been shown to improve the precision as outlined in Sections \ref{sec:magneticExp} and  \ref{sec:waveForm} below. Giblin~\emph{et al.}~\cite{giblin2012} have confirmed under such optimized conditions agreement with the expected quantized value for currents up to 150$\,$pA. The measurement method uses a reference current derived from primary standards. This means it is based on the 1990 units, using the agreed values of the von Klitzing constant $R_{K-90}$ and Josephson constant $K_{J-90}$. The systematic uncertainty of this novel measurement setup has been determined as $1.2 \times 10^{-6}$~\cite{giblin2012}. Improvements for the measurement setup have been discussed in \cite{giblin2014}, reducing the systematic uncertainty so that longer integration times or higher currents would allow an uncertainty reduction. It has been predicted that for a pump delivering $150\,$pA  a standard uncertainty of $6.5\times 10^{-7}$ can be attained after 15 hours of averaging. It should be noted that a QMT experiment with corresponding uncertainty levels will in addition require means of error accounting~\cite{scherer2013, Wulf2013}. Here a combination of serially connected pumps and detectors allows a self-referenced current generation monitoring the pump errors while sourcing the current. It also effectively reduces the uncertainty in the sourced current as information on the error type can be extracted. The progress of this method is reviewed in Section~\ref{sec:erroraccounting}.

Using the above mentioned ULCA the systematic uncertainty is substantially lower, i.e. about $6\times10^{-8}$~\cite{drung2014}. The stable current gain calibrated against primary standards allows direct comparison of the amplified current to a reference current, again traced back to primary standards. At the same time the noise level remains sufficiently low so that within a day of integration, a standard uncertainty of about $1\times10^{-7}$ may be achieved even at only $100\,$pA. Further improvements down to the systematic uncertainty may be achieved using higher currents, e.g., from pumps operated in parallel~\cite{maisi2009,kaestner2010c} or by applying higher pump frequencies.

%%%%%%%%%%%%%%%%%%%%%%%%%%%%%%%%%%%%%%%%%%%%%%%%%%%%%%%%%%%%%%%%
\subsection{Magnetic field effect on quantization accuracy}
\label{sec:magneticExp}

%The effect of magnetic field on quantisation accuracy has been studied so far mostly for GaAs-based tuneable-gate pumps devices.
The single-gate modulated GaAs-based devices described in Sections \ref{sec:plungerBarrier} and \ref{sec:currModel} have been studied extensively after the empirical observations of Wright~\emph{et al.}~\cite{wright2008} and Kaestner~\emph{et al.}~\cite{kaestner2009a} that application of a magnetic field results in plateau enhancement. The accuracy-enhancement effect has been confirmed in several later studies on similar devices~\cite{Leicht2009, kaestner2010d, mirovsky2010, wright2011, giblin2012, Fletcher2012}. For example, for a sine-wave-driven pump at $f = 150\,$MHz the $B$-field was found in \cite{giblin2012} to increase monotonically  the parameter $\delta_2$ (and hence the predicted accuracy, see Section~\ref{sec:currModel}) from about $5$ at $B = 0 \,$ T to $21$ at $B = 14\,$T (see Figure~\ref{fig:GiblinPulse}(d) below). However,
sample-to-sample variation in precision of pumping in lithographically identical structures may be similarly large. For example,  the zero-magnetic-field pump characteristic shown  in Figure~\ref{fig:decayVeri}(a) for sample A corresponds to a plateau lengths value of $\delta_2 = 26.5$ measured at $f=200\,$MHz. This device is lithographically identical to the one in~\cite{kaestner2009a} showing $\delta_2\approx 15$ at $B=3\,$T and a much lower frequency of $f=50\,$MHz. Another zero-magnetic-field high-precision example is the realization of an entirely gate-defined GaAs QD by Seo~\emph{et al.}~\cite{seo2014} operating at $f=500\,$MHz and reaching $\delta_2 = 18.1$. These results suggest that a perpendicular magnetic field is not an absolute requirement for reaching sub-ppm precision levels. Nevertheless, the overall agreement in the literature on the enhancement effect supports the operation in magnetic field environments to increase  the  yield of high-precision GaAs pumps.

Fletcher \emph{et al}.\  \cite{Fletcher2012} have analyzed two mechanisms for  magnetic field influence.
The first mechanism is increased  sensitivity of the tunneling rates to the electrostatic potential (i.e., reduction of effective $\Delta_b$) due to magnetic confinement. Smaller $\Delta_b$ implies  a reduction of $\Delta_{\rm ptb}$ and, consequently, an increase of $\delta_2$ in the framework of the decay cascade model, as explained in Section~\ref{sec:modIntro}. The second mechanism studied in \cite{Fletcher2012} is non-adiabatic excitation of the last electron on the QD to the less-confined higher energy states (see \cite{kataoka2011} and Section~\ref{sec:internalExcitations}) attributed in \cite{Fletcher2012}  to the changing confinement potential shape during the decoupling stage. This  effect has a non-monotonic dependence on the magnetic field resulting in a range of fields where the excitations are most pronounced. For the devices studied in \cite{Fletcher2012} the excitations are significantly weakened at $B > 12\,$T. 
Note however, that stronger $B$-fields require the modulation amplitude to be increased \cite{wright2011} which in turn has been shown to potentially degrade the precision~\cite{YamahataPRB2014}. The highest $B$-field of $B = 30\,$T applied to pump operation did not, however, show a plateau enhancement beyond 15$\,$ T~\cite{kaestner2010d}.

In summary, the increased sensitivity of the tunneling rate to the electrostatic potential and the suppression of non-adiabatic excitations at high magnetic fields are the most likely reasons for the observed enhancement in accuracy. However, the delicate interplay of magnetic and electrostatic confinement %analyzed by Fletcher \emph{et al}.\ \cite{Fletcher2012} 
during  the QD decoupling stage
 may also be responsible for the realization- and device-specific manifestation of this effect.

%This would be another indication for rate determined quantization as assumed e.g. in the decay cascade model.

%\begin{figure}
%\includegraphics{fig_bEnhancement.pdf}
%\caption{
%\label{fig:bEnhancement}}
%\end{figure}

%For the device geometry employed by Fletcher~\emph{et al.}~\cite{Fletcher2012} roughly a monotonic increase in the plateau length measured by the decay cascade $\delta$-paramter has been observed up to 14$\,$T, as shown in Fig.~\ref{fig:bEnhancement}. For devices in Fig.~\ref{fig:B_LL}(a) it was found that beyond 5$\,$T no significant improvement occurs.

%\begin{figure}
%\includegraphics{fig_B_LL.pdf}
%\caption{LL
%\label{fig:B_LL}}
%\end{figure}

%		even-odd (multiple particle feature)
%		Explanation: Fletcher -- enhancement of sensitivity of tunnel rates to barrier potential;
%		However, B-field may also reduce plateau length in certain QD structures: DQD like, see Fig.~\ref{fig:B_DQD}

%\begin{figure}
%\includegraphics{fig_B_DQD.pdf}
%\caption{Even Odd
%\label{fig:B_DQD}}
%\end{figure}

%		If leads are brought into QHE regime the pump with parabolic channel shape shows plateau structure shifts correlated with LL crossing 		see Figure~\ref{fig:B_LL};

%%%%%%%%%%%%%%%%%%%%%%%%%%%%%%%%%%%%%%%%%%%%%%%%%%%%%%%%%%%%%%%%%%%%%%%%%%%%%%%%%%%%%%%%5
\subsection{Degrading effect of internal excitations}
\label{sec:internalExcitations}

\begin{figure}
\includegraphics{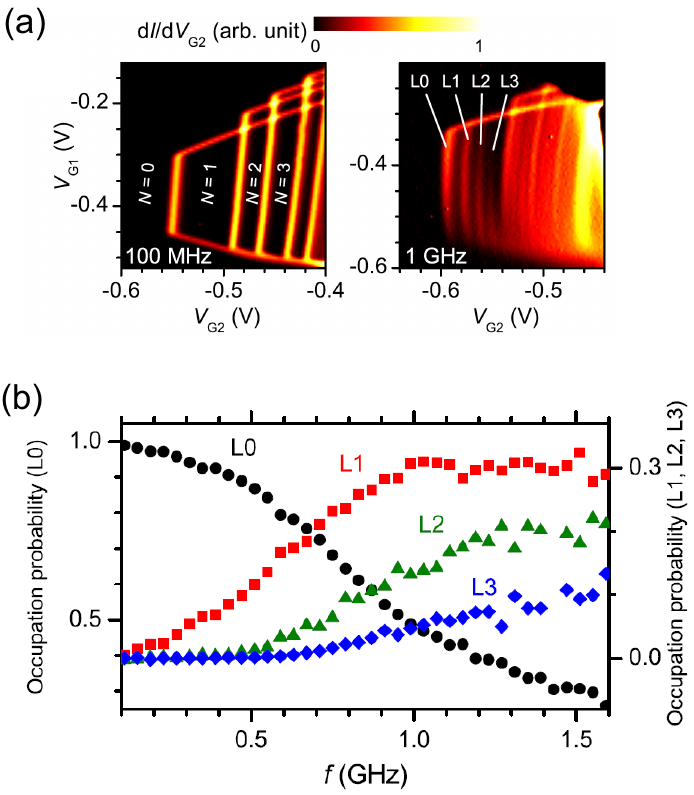}
\caption{(a) Derivative of current $dI/dV_{G2}$ as function of $V_{G1}$ and $V_{G2}$, taken with $f=100\,$MHz and $1\,$GHz at $B = 6\,$T. (b) Extracted occupation probability of the excited states for $B = 6\,$T. Reprinted with permission from \cite{kataoka2011}. Copyright 2011 by the American Physicl Society.
\label{fig:excitedStatesKataoka}}
\end{figure}

Deviations from adiabaticity in the distribution of electronic energy 
becomes a challenge for  quantized pumping as the  repetition frequency or the measurement precision is increased.
Energy dissipation in components with a continuous, metal-like spectrum (such as source and drain leads) is expected to raise  the local effective temperature. This will degrade the plateaus if the dominating cause of error is thermal (i.e., traceable to the smearing of the Fermi level in \eqref{eq:detailedW} or its equivalent). Although local heating has been discussed in \cite{Janssen2001,Utko2006,Chan2011a}, systematic studies of the temperature effect in gate-driven tunable-barrier pumps have not yet been reported to the best of our knowledge.
Experimentally, it is easier to link degradation of plateaux quality to non-adiabatic excitation of energy levels if the latter are discrete.  A short overview of possible physical mechanisms for non-adiabatic excitation during charge capture has been given at the end of Section~\ref{sec:modIntro}; here we discuss the available experimental evidence of individual excited states in the pump  characteristics.

%Such internal excitations may be caused, for instance, by preferential loading of electrons into excited states~
%\cite{fricke2011} or by exciting electrons experiencing the fast potential variation during the decoupling process
%\cite{Robinson2001, kataoka2011}. Whether these excitations show up in the capturing statistics depends on the relation
%between the relaxation time and the timescale for decoupling. The actual time for this transition is a fraction of the period. 
%For a given pump frequency it may be tuned using pulse shape variation as discussed in Section~\ref{sec:waveForm}. If sufficient time is allowed for the decoupling process then excitations may relax before they have an effect on the backtunneling and consequently the current. 
%Typically, the tunneling rate of the excited state will be larger than that of the ground state. Hence, the pumped current will be reduced which can be used for probing excited states. 
%Note that fast potential deformations may also cause excitations in the leads, as discussed in Section~\ref{sec:nonad}.

\begin{figure}
\includegraphics{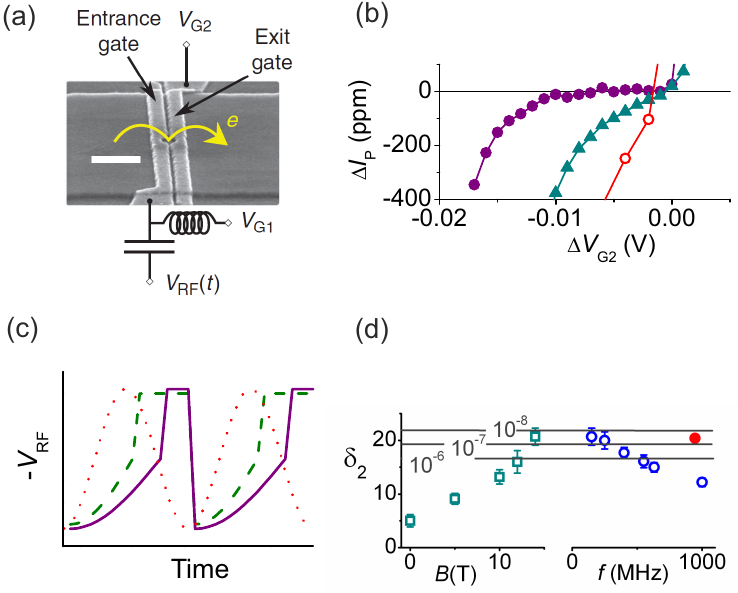}
\caption{Effect of different waveforms, from Giblin~\emph{et al.}~\cite{giblin2012}. (a) Scanning electron micrograph of the device. The scale bar indicates $1 \, \mu$m. (b) Fractional deviation of pump currents from $q_ef$ in ppm, as a function of normalized exit barrier defining voltage at 630$\,$MHz, using sine wave drive (open circles), waveform AWG1 (closed triangles) and waveform AWG2 (closed circles). (c) Corresponding waveforms to obtain data in (b), with sine wave, AWG1 and AWG2 shown as dotted, dashed, and solid line, respectively. (d) Fitting parameter $\delta_2$ as function of magnetic field (for $f=150\,$MHz sine wave drive) and frequency $f$. Horizontal lines show the value for $\delta_2$ where the decay cascade model predicts deviations from the quantized value of 1, 0.1 and 0.01 ppm. The closed red circle indicates the AWG drive. Reprinted by permission from Macmillan Publishers Ltd: Nat. Commun. \cite{giblin2012}, copyright 2012.}
\label{fig:GiblinPulse}
\end{figure}

Kataoka~\emph{et al.}~\cite{kataoka2011} have observed discrete  excitations in a  QD operated as a single-gate capture-limited pump (see  Section~\ref{sec:plungerBarrier}) where the dot is defined by specifically shaped gates in a relatively wide channel ($2\,\mu$m), similar to the device shown in Fig.~\ref{fig:GiblinPulse}(a). The confinement potential transverse to the transport direction is created by the field effect of the gates. Figure~\ref{fig:excitedStatesKataoka}(a) shows the derivative of the pump current $dI/dV_{G2}$ as function of the dc components of the gate voltages $V_{G1}$ and $V_{G2}$ that define the QD;  large-amplitude sinusoidal  modulation is
applied to $V_{G1}$. For high-frequency modulation the plateaus develop a strong slope with a series of miniature plateaus separated by the peaks in the derivate (L0, L1, L2, L3). This structure has been interpreted in \cite{kataoka2011} as a statistic superposition of charge capture probabilities conditioned on the electron being in a specific quantum state.
For example, tuning of $V_{G2}$ to the mini-plateaux between L0 and L1 results in charge capture only if the electron
is in the ground state. If the model of gradual decoupling described in Section \ref{sec:modIntro} is applied to two  single-electron states that differ in energy by $\Delta \epsilon$, the expected separation in integrated escape rates is $X_1^{L(i+1)}/X_1^{L(i)}= \exp (\Delta \epsilon/\Delta_b +  \Delta \epsilon/\Delta_{\rm ptb})$. Such separation corresponds (see \eqref{eq:fittingansatz} and accompanying discussion) to a gate-voltage difference between the ministeps $L(i)$ and $L(i+1)$ that is proportional to $\Delta \epsilon$. The correlation of peak positions ($L1$-$L3$ features) with  excited state energies of a single electron confined by a 2D parabolic potential and a perpendicular magnetic field (the Fock-Darwin spectrum  \cite{kouwenhoven2001}) has been  established experimentally as function of $B$ between $0$ and $10 \, \mathrm{T}$ \cite{kataoka2011}.

The occupation probability for the different excited states has been deduced from the current value on the mini-plateaus, the results from \cite{kataoka2011}  are shown in Fig.~\ref{fig:excitedStatesKataoka}(b). At high frequencies a population inversion occurs which makes thermal excitation an unlikely origin. Further exploration of the excitation mechanism in similar devices has been conducted in \cite{Fletcher2012} where suppression of excitations at very high magnetic fields was found (see previous section).  Relatively high rf power employed in the experiments described above compared to studies of single-gate pumping in narrow-channel GaAs-based devices \cite{kaestner2008, Hohls2012} 
may be a contributing factor to the good visibility of the excited-states pattern which has not been seen in the latter studies. 

A similar spectrum of excited states for a single captured electron has been observed in SAW-driven GaAs pumps by He \emph{et al.}~\cite{he2014}. They report greater sensitivity to magnetic field attributed to elliptic shape of the QD as inferred from comparing the evolution of the excitation features with the $B$-field to the corresponding generalization of the Fock-Darwin spectrum. 

The first excited state in a Si-based tunable QD has been identified in a pump current characteristic in the experiments by Rossi \emph{et al.\ }~\cite{Rossi2014}. The distance between features corresponding to the ground and the excited states respectively was found to grow with increasing electrostatic confinement, similar to the effect of magnetic confinement on the Fock-Darwin spectrum of a GaAs-based QD.

Robustness of the excitation pattern in Fig.~\ref{fig:excitedStatesKataoka}(a) to changes in the dc component on the modulated gate $V_{G1}$ suggests that the effectiveness of the relevant excitation mechanism does not depend on the time the dot spends well-coupled to the source (i.e.\ independent of the duration of the loading stage (i) in terms of Figure \ref{fig:basicPrinciple} from Section \ref{sec:currModel}). Thus the excitation process is likely to be associated with the decoupling stage. A different type of fine structure in pump characteristics has been found by 
Fricke \emph{et al.}~\cite{fricke2011} who report signatures of the excited states in a single-gate pump operating close to the loading-limited regime (corresponding to line $C$ in the schematics of Figure \ref{fig:basicPrinciple}(b)).
These signatures could be more easily identified in the charge level of a node between two pumps operated in series and locked to the same output level via the mechanism discussed in Section~\ref{sec:erroraccounting}. The absence of a resolved fine structure in the current for the loading-defined excitation mechanism has been attributed to efficient relaxation to the ground state before the decoupling stage commences \cite{fricke2011}.
%\commVK{Comment on narrow channels devices: power of less than -15 dBm for quantisation in B of 5 to 9 Tesla, and below %-20 dBm for $B=0$ quantisation, compared to  between 0 to 5 dBm through a 6 dB attenuator, as compared to a total power %of -25 dBm in narrow-channel devices~\cite{kaestner2008}. For 0 dBm the corresponding peak-to-peak amplitude of gate %modulation was estimated as 600$\,$mV~\cite{kataoka2011}.} 

The variety of device-dependent manifestations of the single-particle spectrum in strongly modulated tunable-barrier QDs seems to limit at the present stage any prospects for a simple conclusion regarding the effect of internal excitations on the accuracy of the quantized current.

%%%%%%%%%%%%%%%%%%%%%%%%%%%%%%%%%%%%%%%%%%%%%%%%%%%%%%%%%%%%%%%%%%%%%%%%%%%%%%%%%%%%%%%%%%%%%%%%
\subsection{Optimizing the driving waveform}
\label{sec:waveForm}

Different phases of a pumping cycle in a non-adiabatic pump require different coupling conditions between the source, the QD and the drain. These conditions can be optimised for speed and accuracy by adjusting the driving wave-form.
A significant improvement in capturing precision for the class of single-gate modulated devices discussed in Section \ref{chap:quantModel} has been obtained by Giblin~\emph{et al.}~\cite{giblin2012} by dividing the waveform into a ``slow'' and a ``fast'' part. The slow part of the waveform with repetition frequency $f$ is approximately a segment of the sine wave of frequency $f/5$. 
This slows down the loading and the capture processes (see phases (i) and (ii) in Figure~\ref{fig:basicPrinciple}) compared to the transfer and emission stages (phases (iii) and (iv), respectively) while keeping the repetition frequency high.
 This design has been motivated by the empirical finding that an increase in frequency reduces the quantization quality, as shown in Fig.~\ref{fig:GiblinPulse}(d). The driving waveforms were applied to the entrance gate of a QD defined by the specifically shaped gates in a relatively wide channel shown in Figure~\ref{fig:GiblinPulse}(a). Figure~\ref{fig:GiblinPulse}(b) shows the pump current $I_p$ in ppm fractional deviation from $q_e f$ as function of the normalised voltage $\Delta V_{G2}$ controlling  the barrier of the dot over which the electron is emitted. There modulation waveforms $V_{RF}$  shown in Figure~\ref{fig:GiblinPulse}(c) were applied. When driven by a sine wave (dotted line in (c)) the plateau at $f = 630\,$MHz almost disappears (open circles in (b)). The plateau structure at the same frequency becomes progressively better for waveforms labeled AWG1 (dashed line in (c)) and AWG2 (solid line in (c)), respectively, reaching metrologically relevant levels. 

In single-gate Si-based tuneable barrier pumps, pulsed modulation has been investigated \cite{fujiwara2008,miyamoto2008,Yamahata2014,YamahataPRB2014}.  We have already discussed the robustness of the capture process to the pulse rise time in Section \ref{sec:currModel} (see Figure \ref{fig:Fujiwara} and the accompanying discussion). Miyamoto~\emph{et al.}~\cite{miyamoto2008} investigated the influence of the pulse shape on the emission dynamics. This has been achieved by first tuning the pump into the capture-dominated regime and reading off the number of the periodically captured electrons from $\qav{n}=I/(q_e f)$. Subsequently, emission pulse height $V_{G1L}$ at the pump gate and its hold time $t_{G1L}$ are varied. The resulting current $I$ is then a measure for the number of electrons emitted to drain, $\qav{n_e}= I/(q_e f)$. The results for three captured electrons have been fitted to a time-dependent solution of the master equation for the probability $P_m$ for $m$ electrons remaining on the dot at the end of the emission pulse: $dP_m/dt = P_{m+1}/\tau_n - P_m/\tau_{n+1}$ under the conditions $n+m =3$ and $P_3(0)=1$, which is equivalent to our \eqref{eq:det:kin1} with $\Gamma_n = \tau_{3-n}^{-1}$ and $f(\mu_n) \to 0$ (emission only) for $t=0 \ldots t_{G1L}$~\cite{Kashcheyevs2014}. The fit parameters $\tau_n$ give direct information on the gate voltage dependence of the emission rate $\Gamma_n(V_{G1L})$. The temperature dependence of the inferred $\Gamma_n(V_{G1L})$  in the range of $T=16 \rm{K} \ldots 28 \rm{K}$ and similar tendencies in the subthreshold regime of the MOSFET corresponding to the same barrier are indicative of thermally activated hopping. In this experiment the range for hold times is limited by the resolution of the current measurement. Further expansion of the range would require a direct counting technique of the type employed in \cite{yamahata2011,YamahataPRB2014,Fricke2013} and described in Section \ref{sec:CountStatModel}. In a  fashion similar to the methods of \cite{miyamoto2008}, both the capturing and emission rates have been determined by Yamahata~\emph{et al.} \cite{Yamahata2014}. The results show that by analyzing the emission and capturing statistics in pulse-driven devices one can optimise selectively the relevant parts of the cycle, namely loading and emission. The corresponding hold times may be increased only until the desired precision is achieved, in order not to sacrifice the repetition rate.

\begin{figure}
\includegraphics{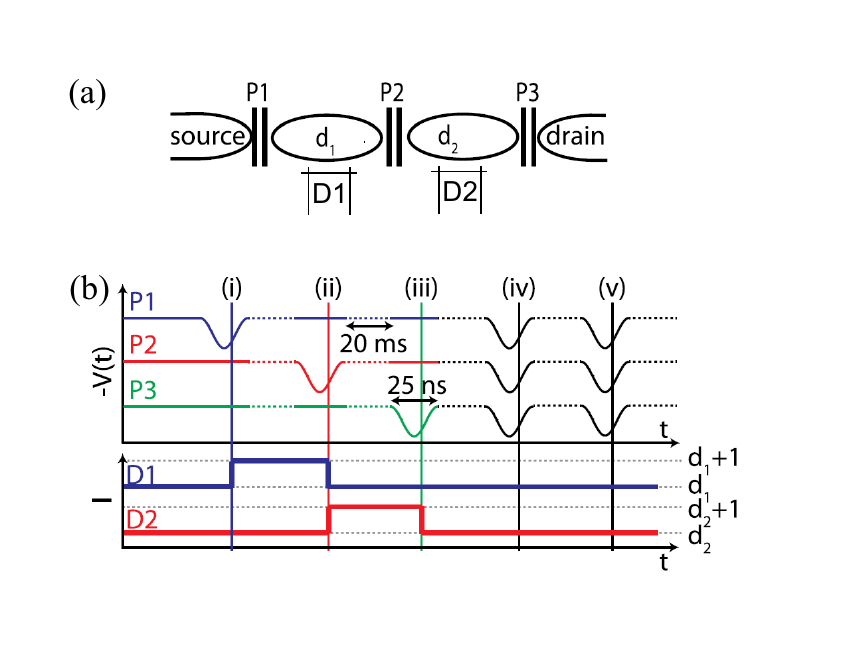}
\caption{(a) Schematic of error accounting circuit consisting of three pumps connected by two nodes with charge configuration ($d_1$, $d_2$) which is monitored by single-charge detectors D1 and D2. (b) Pulse sequence $V(t)$ applied to the pumps, triggering single-electron transfer events. Traces marked D1 and D2 show nominal detector signals for error-free pumping. Adapted from Fricke \emph{et al.}~\cite{fricke2014}. 
\label{fig:counting_device}}
\end{figure}

%%%%%%%%%%%%%%%%%%%%%%%%%%%%%%%%%%%%%%%%%%%%%%%%%%%%%%%%%%%%%%%%
%%%%%%%%%%%%%%%%%%%%%%%%%%%%%%%%%%%%%%%%%%%%%%%%%%%%%%%%%%%%%%%%
%%%%%%%%%%%%%%%%%%%%%%%%%%%%%%%%%%%%%%%%%%%%%%%%%%%%%%%%%%%%%%%%

\subsection{Error accounting}
\label{sec:erroraccounting}

\begin{figure*}
\includegraphics[width=\largefig]{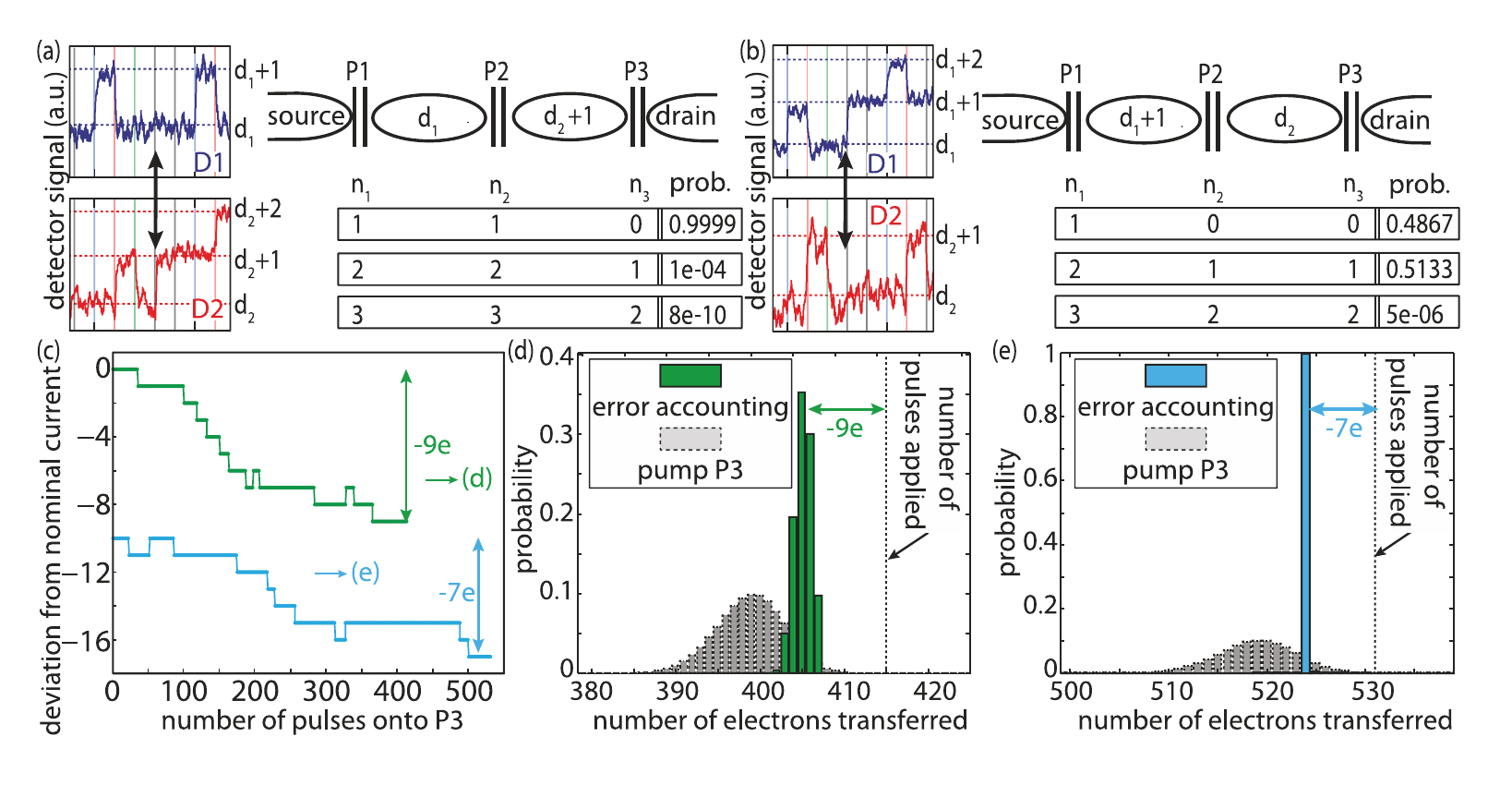}
\caption{Analysis of pumping errors by charge detection, adapted from Fricke \emph{et al.}~\cite{fricke2014}. Measured detector signal time traces showing error events  (arrow) in (a) and (b). Different error scenarios with corresponding probabilities are shown in the table. Panel (c) shows the deviation from nominal current across P3 versus the pulse index for two working points of the pumps. (d) Probability distribution of the number of pumped electrons of the green (upper) trace from (c) compared to the case without error accounting for P3. (e) Same plot as (d) but for the blue (lower) trace from (c).
\label{fig:counting_data}}
\end{figure*}

The reliability of any single-electron current source is eventually limited by the stochastic nature of the underlying quantum mechanical tunneling process, whereas typically its relative error rate increases with current output. The counting of electrons passing randomly through a nanostructure has been explored~\cite{bylander1, fujisawa2006}, however, the sensitivity and bandwidth of available detectors limit current output and uncertainty. The method of error accounting proposed by Michael Wulf~\cite{Wulf2013} offers a way to overcome this issue. Here a combination of serially connected pumps and detectors is used such that during operation of the pumps the \emph{rare} stochastic errors are detected to determine the deviations from the nominal current $I=q_ef$. This would  allow in principle to increase the current output and account for a correspondingly increased error rate.

The challenge of serial operation is that a slight mismatch in the pump current would lead to charge built-up on the nodes, which in turn leads to shifts in the electrochemical potential at the entrance and exit of the individual pumps and thereby possibly affecting their quantization. A feedback circuit as introduced by Keller \emph{et al.}~\cite{keller1999} may be employed to keep the voltage across the pump near zero. This scheme has been successfully demonstrated for a device which pumps single charges onto a node acting as capacitance standard~\cite{keller1999, keller2007a}. To the best of our knowledge this feedback scheme has not yet been realized for serial operation. Serial operation of two single-gate modulated tunable barrier devices (similar to the one in Figure~\ref{fig:decayVeri2}(a)) has been investigated by Fricke \emph{et al.}~\cite{fricke2011}. The two pumps have been linked by a few-micron-wide mesoscopic island. Despite the unavoidable charge pile-up, stable serial operation of the two pumps has been demonstrated where the \emph{second} pump can be locked onto the quantized current of the \emph{first} one by a feedback due to charging of the mesoscopic island. This \emph{hierarchical locking} greatly facilitates the search for stable serial operation avoiding an external feedback circuit.

A low frequency version of a full error accounting circuit has recently been demonstrated by Fricke~\emph{et al.}~\cite{fricke2014}. The work employed three single-gate operated tunable barrier pumps (P1, P2, P3) connected in series, as shown schematically in Fig.~\ref{fig:counting_device}(a). Changes in the charge configuration ($d_1$, $d_2$) of the intermediate nodes are monitored by the two aluminum based SETs~\cite{Fulton1987, kuzmin1989}, D1 and D2, operated at fixed bias voltages.
% The current, acting as detector signal, is sampled at $12\,$kS/s, which is about 20 times faster than the detector response.

Pumping through this hybrid structure is induced by pulses as shown in Fig.~\ref{fig:counting_device}(b). First each pump is triggered once for calibration of detectors D1 and D2 and determination of the error rates of the individual pumps. The last two pulses trigger all pumps synchronously for the actual error accounting, where the detector signal should not change, as shown at the bottom of Fig.~\ref{fig:counting_device}(b). If it does not stay constant, an error event has occurred. An example is shown in the real detector trace of Fig.~\ref{fig:counting_data}(a), where D2 indicates an additional electron on node 2. To identify the error for this particular case and \emph{correct} the corresponding current output three possible scenarios should be considered: (1) P3 failed to pump one electron, (2) P1 and P2 each pump one additional electron, or (3) they pump two additional electrons and P3 one additional electron. From the characterization during the marker-sequence corresponding probabilities have been derived, as shown in the table in Fig.~\ref{fig:counting_data}(a), and hence, P3 missing a cycle is the most likely error event with a probability of 0.9999. The situation for other error events may not be as certain. An example is shown in Fig.~\ref{fig:counting_data}(b).

By statistical analysis of long series of pump events one obtains the output current and its uncertainty. The actual current corresponds to the number of electrons transferred across P3 to drain. Fig~\ref{fig:counting_data}(c) shows two example traces. After each trace a probability distribution can be obtained for the number of electrons transfered through the whole hybrid structure. Fig~\ref{fig:counting_data}(d) compares the uncertainty of P3 with the accounted transfer for the green trace in Fig~\ref{fig:counting_data}(c). A reduction in uncertainty can be seen (variance of probability distribution), together with the corrected number of transferred electrons.

By choosing the working points so that only well distinguishable errors occur, the probability distribution can be made very narrow. This situation corresponds to the blue trace in Fig.~\ref{fig:counting_data}(c) and the distribution in (e). After error accounting the uncertainty is reduced by a factor of 50. Note that tuning pumps to asymmetric transfer rates makes transfer of more than one electron very unlikely. Well distinguishable errors can thus be obtains. Note further that for pumps operated in the decay cascade regime an asymmetry is already present at the working point of highest precision (see Section~\ref{sec:currModel}).

The above work has demonstrated the principle of error accounting albeit at a very low repetition frequency of about $30\,$Hz. The integration of currently available rf-SET detectors~\cite{schoelkopf1998} with an increased bandwidth of 50 kHz has been analyzed in~\cite{fricke2014}. At a pump frequency of 1 GHz and a pump error probability of $1 \times 10^{-6}$ a corresponding error rate in the kHz range may be detected with this technology. Five serial pumps would allow self-referenced GHz-operation of already demonstrated tunable barrier pumps at very low relative uncertainty of less than $10^{-8}$. Issues to be solved for this ambitious goal are cross-talk between detectors and pumps as well as managing increased circuit complexity.

\section{Conclusions and outlook}
\label{sec:outlook}

Enhanced charging energy of sufficiently small structures has proved the access route to the
discreteness of charge since the pioneering discovery of Millikan~\cite{Millikan1913} more than a century ago.
Modern semiconductor nanotechnology enables controlled charging of the quantization-defining element (e.g., the quantum dot) through tunable barriers. High-fidelity single-electron manipulation requires good on-off switching
of these barriers over many orders of magnitude in conductance. Achieving such modulation with the field-effect implies large amplitude changes of electrostatic potential in the vicinity of the quantum dot.
It was the development of efficient non-adiabatic pumping schemes which 
are free from the requirement to navigate the energy landscape with single-electron precision that has enabled robust and accurate operation of charge pumps at high frequencies.
The focus of modeling and device optimization has shifted from the whole pumping cycle to separate stages
(loading, decoupling, isolation, and emission) with clearer trade-off requirements for each stage.

A strong motivation for the field remains the potential application of an accurate quantized current source in metrology. Two directions of active development can be identified. On the one hand, an error probability of $10^{-6}$ at GHz pump frequencies has been experimentally demonstrated~\cite{giblin2012} and the work of improving precision and operating frequency of individual pumps is ongoing. New strategies include enhanced tuning of the electrostatic confinement~\cite{Rossi2014}, exploitation of charge traps as the quantization-defining elements \cite{Yamahata2014}, utilization of industrial nanoelectronics fabrication processes \cite{jehl2013}, and control  of quantization-degrading energy scales with optimal driving wave-forms (Sections \ref{sec:modIntro} and \ref{sec:currModel}). On the other hand, the techniques of error accounting (Section~\ref{sec:erroraccounting}) are being developed with the aim of measuring directly the deviation of the output current from the nominal quantized value. A circuit with five pumps in series, each performing  at the present record  precision level and integrated with the currently available detector technology, has been predicted~\cite{fricke2014} to reduce the relative uncertainty below the target level of $10^{-8}$. With further improvement of individual pumps, the same error accounting target could be achieved with fewer pumps, thus reducing the circuit complexity. An ambitious goal of such self-referenced operation is a direct closure of the quantum metrological triangle that would test the primary assumptions of electrical quantum metrology (Section~\ref{sec:metro}).

Promising applications of the single-electron manipulation technology already emerge in  domains outside of metrology.
One of such areas is few-body mesoscopic solid-state physics in high mobility semiconductor structures. The tunable-barrier QD has been employed as a highly-tunable on-demand source of ballistic electrons and electron pairs for solid-state electron quantum optics experiments~\cite{FletcherPRL2013,NielsNature2014}. 
Opportunities for wave-packet shaping and coherence of ballistic electrons at high energies of tens of meV above the Fermi energy are yet to be explored compared  to the relatively well-studied sub-meV regime of Hall edge electron quantum optics \cite{Bocquillon2014}.

One of the ambitious goals for mastering the active switching of localized and propagating single-electron modes at the quantum level is the development of a ``flying qubit'' \cite{Hermelin2011,McNeil2011} which would provide the necessary entanglement bus between stationary qubits in a semiconductor architecture for quantum information processing~\cite{MortonNature2011}. Another active area of research which stands to benefit from high-fidelity control of individual transport events is statistical mechanics and thermodynamics with single-particle resolution~\cite{kueng2012, saira2012, pekola2015}, with strong interest in testing fundamental non-equilibrium fluctuation theorems \cite{RMP2011}.

\begin{acknowledgements}
We thank  Dietmar Drung, Lukas Fricke,  Akira Fujiwara, Stephen Giblin, Frank Hohls, Jan Theodor Janssen, Xavier Jehl, Christoph Leicht, Alessandro Rossi, Hansj\"org Scherer, Hans Werner Schumacher, Janis Timoshenko and Gento Yamahata for discussions and feedback on various parts of the manuscript. 
This work has been supported by EC FP7 under Project SiAM No.\ 610637.
V.K. has also  been supported by Latvian Council of Science (grant no.\ 146/2012).
\end{acknowledgements}
% Uncomment for ROP
% \section*{References}

%\bibliography{References} 

\end{document}